\documentclass[aps,pre,reprint,noshowpacs,superscriptaddress,twocolumn,notitlepage]{revtex4-1}

\usepackage{graphicx}
\usepackage{amsmath,amsfonts,amsthm}
\usepackage{amssymb}

\usepackage[colorlinks = true,
            linkcolor = teal,
            urlcolor  = teal,
            citecolor = teal]{hyperref}
\usepackage{mathtools}
\usepackage{mathrsfs} 
\usepackage{bm}
\usepackage{xcolor}
\usepackage[caption=false]{subfig}
\usepackage[utf8]{inputenc}
\usepackage[T1]{fontenc}
\usepackage{braket}

\usepackage[normalem]{ulem}

\newtheorem{definition}{Definition}

\begin{document}
\title{Network comparison and the within-ensemble graph distance}

\author{Harrison Hartle}
\affiliation{Network Science Institute, Northeastern University, Boston, MA, USA}
\author{Brennan Klein}\thanks{correspondence: klein.br@northeastern.edu}
\affiliation{Network Science Institute, Northeastern University, Boston, MA, USA}
\affiliation{Laboratory for the Modeling of Biological and Socio-Technical Systems, Northeastern University, Boston, MA, USA}
\author{Stefan McCabe}
\affiliation{Network Science Institute, Northeastern University, Boston, MA, USA}
\author{Alexander Daniels}
\affiliation{Vermont Complex Systems Center, University of Vermont, Burlington, VT, USA}
\author{Guillaume St-Onge}
\affiliation{D\'epartement de Physique, de G\'enie Physique et d'Optique, Universit\'e Laval, Qu\'ebec, Canada}
\affiliation{Centre Interdisciplinaire de Modélisation Mathématique, Université Laval, Québec, Canada}
\author{Charles Murphy}
\affiliation{D\'epartement de Physique, de G\'enie Physique et d'Optique, Universit\'e Laval, Qu\'ebec, Canada}
\affiliation{Centre Interdisciplinaire de Modélisation Mathématique, Université Laval, Québec, Canada}
\author{Laurent Hébert-Dufresne}
\affiliation{Vermont Complex Systems Center, University of Vermont, Burlington, VT, USA}
\affiliation{D\'epartement de Physique, de G\'enie Physique et d'Optique, Universit\'e Laval, Qu\'ebec, Canada}
\affiliation{Department of Computer Science, University of Vermont, Burlington, VT, USA}

\date{August 5, 2020}

\begin{abstract}
Quantifying the differences between networks is a challenging and ever-present problem in network science. In recent years a multitude of diverse, ad hoc solutions to this problem have been introduced. Here we propose that simple and well-understood ensembles of random networks---such as Erd\H{o}s-R\'{e}nyi graphs, random geometric graphs, Watts-Strogatz graphs, the configuration model, and preferential attachment networks---are natural benchmarks for network comparison methods. Moreover, we show that the expected distance between two networks independently sampled from a generative model is a useful property that encapsulates many key features of that model. To illustrate our results, we calculate this \textit{within-ensemble graph distance} and related quantities for classic network models (and several parameterizations thereof) using 20 distance measures commonly used to compare graphs. The within-ensemble graph distance provides a new framework for developers of graph distances to better understand their creations and for practitioners to better choose an appropriate tool for their particular task.
\end{abstract}

\maketitle

\section{Introduction}\label{sec:intro}
Quantifying the extent to which two finite graphs structurally differ from one another is a common, important problem in the study of networks. We see attempts to quantify the dissimilarity of graphs in both theoretical and applied contexts, ranging from the comparison of social networks \cite{berlingerio2012netsimile, koutra2016deltacon, Tsitsulin2018}, to time-evolving networks \cite{Bagrow2019, Donnat2018, Masuda2019, Torres2019, Monnig2018}, biological networks \cite{Donnat2018}, power grids and infrastructure networks \cite{Schieber2017}, object recognition \cite{wilson2008study}, video indexing \cite{bunke1998graph}, and much more. Together, these network comparison studies all seek to define a notion of dissimilarity or \textit{distance} between two networks and to then use such a measure to gain insights about the networks in question.

However, it is often unclear which network features a given graph distance will or will not capture. For this reason, rigorous benchmarks must be established in order to better understand the tendencies and biases of these distances. We adopt the perspective that random graph ensembles are the appropriate tool to achieve this task. Specifically, by sampling pairs of graphs from within a given random ensemble with the same parameterization and measuring the graph distance between them, we create a benchmark that allows us to better understand the sensitivity of a given graph distance to known statistical features of an ensemble. Ultimately, a good benchmark would characterize the behavior of graph distances between graphs sampled from both \textit{within} an ensemble and \textit{between} different ensembles. We tackle the former in this paper, noting a rich diversity of behaviors among commonly used graph distance measures. Even though this work focuses on within-ensemble graph distances, these results guide our understanding of how any two sets of networks structurally differ from each other regardless of if those sets are generated by the same random ensemble or another network-generating process. Put simply, the approach introduced in this work is general and can be used to develop a number of graph distance benchmarks.

There are many approaches used to quantify the dissimilarity between two graphs, and we highlight 20 different ones here. Given the large number of algorithms considered in this work, we find it useful to systematically characterize each of these measures. We do so by breaking them down into ``description-distance'' pairs. That is, every graph distance measure can be thought of as 1) computing some \textit{description} or property of two graphs and 2) quantifying the difference between those descriptions using some \textit{distance} metric.

\subsection{Formalism of Graph Distances}\label{sec:formalism}
\subsubsection*{Graph Descriptors}

\begin{definition} 
    A \textit{graph description} $\Psi$ is a mapping from a set of graphs $\mathcal{G}$ to a space $\mathcal{D}$,
    \begin{equation}
    \Psi : \mathcal{G} \rightarrow \mathcal{D}.
    \end{equation}
\end{definition}

The set $\mathcal{G}$ is that of all finite labeled simple graphs, and the space $\mathcal{D}$ is known as the {\it graph descriptor space}. Typically, $\mathcal{D}$ is $\mathbb{R}^{ l\times m}$ for integers $l,m$ or is a space of probability distributions. Given a description $\Psi$, the \textit{descriptor} of graph $G$, denoted $\psi_{G}$, is the element of $\mathcal{D}$ to which $G$ is mapped; $\psi_G=\Psi(G)$.

\subsubsection*{Descriptor Distances}

\begin{definition}
A distance maps a pair of descriptors to a nonnegative real value,

\begin{equation}
d: \mathcal{D} \times\mathcal{D} \rightarrow \mathbb{R}_{+}
\end{equation}
and satisfies the following properties for all $x,y\in\mathcal{D}$: 
\begin{enumerate}
    \item $d(x,y) = d(y,x)$ (Symmetry)
    \item $d(x,x) = 0$ (Identity Law)
\end{enumerate}
\end{definition}

The properties listed in this definition are general, and they do not restrict the large possibility of measures we might use, while also providing a clean separation between how we choose to describe graphs and how we calculate the differences between those descriptions. A common property when considering distance measures is the \textit{triangle inequality}; however we have not included this in the list above as not all commonly used graph distances obey this property \cite{Bento2019}. As in the case of pseudometrics, $d(x,y)=0$ does not always imply $x=y$ \cite{Torres2019} \footnote{For example, two cospectral but non-identical graphs would have distance zero according to any spectral distance measure.}.

\subsubsection*{Graph Distances}
\begin{definition}
Given a set of graphs $\mathcal{M}\subseteq\mathcal{G}$, a graph description $\Psi$, its descriptor space $\mathcal{D}$, and a distance $d$ on $\mathcal{D}$, the associated graph distance measure $D:\mathcal{M}\times\mathcal{M}\rightarrow\mathbb{R}_+$ is a function defined by
\begin{equation}
    D(G,G')=d(\psi_G,\psi_{G'}).
\end{equation}
\end{definition}

Every graph distance quantifies some notion of dissimilarity between two graphs \footnote{Throughout this paper, we use the term ``graph distance'' or ``distance'' to refer to a dissimilarity measure between two graphs satisfying the properties we detail in Section \ref{sec:formalism}. This language is somewhat imprecise from a mathematical perspective; many graph distances do not meet all the criteria of distance \textit{metrics}. We have chosen to keep the term ``graph distance'' at the cost of some informality to maintain consistency with much of the existing literature we draw upon. We thank an anonymous reviewer for their help in clarifying this matter.}.

\subsubsection*{Network spaces}
\begin{definition}
Given a distance $d$ and description $\Psi$ on descriptor space $\mathcal{D}$ and a set of graphs $\mathcal{M}\subseteq{\mathcal{G}}$, the associated network space, denoted $(d,\Psi,\mathcal{M})$, is the set of descriptors mapped to by $\Psi$ from graphs in $\mathcal{M}$, equipped with $d$ as a distance measure.
\end{definition}

The network space $\left(d, \Psi, \mathcal{M} \right)$ consists of $|\mathcal{M}|$ points in $\mathcal{D}$, namely $\left\{\psi_G\right\}_{G\in\mathcal{M}}\subseteq \mathcal{D}$---giving rise to $|\mathcal{M}|(|\mathcal{M}|+1)/2$ distance values, one for each pair of descriptions of elements of $\mathcal{M}$. 

Fundamental questions naturally arise. Does a network space capture known properties of a given ensemble of graphs? This question we can begin to answer by considering sets of graphs with known properties: i.e., random graph models.

\subsubsection*{Models}
\begin{definition}
A model $M_{\vec{\alpha}}$ is a process which generates a probability distribution $\mathbb{P}_{\vec{\alpha}}$ over a set of graphs $\mathcal{M} \subseteq \mathcal{G}$, where $\vec{\alpha}$ is a vector of parameters needed by the model to generate the distribution.
\end{definition}

Models are typically stochastic processes that take some parameters as inputs and generate sets of graphs. The probability distribution of model $M_{\vec{\alpha}}$ is then defined over the set of graphs that have non-zero probability of being generated given the model and its parameters $\vec{\alpha}$. For many well-known models, we have a deep understanding of how the structure of sampled graphs is influenced by the parameter values. Using our knowledge of how parameters affect graph structure, we can see how well the expected features of a given model are reflected by the structure of each network space.

\subsection{This study}
Herein, we apply a variety of graph distances to pairs of independently and identically sampled networks from a variety of random network models, over a range of parameter values for each, and consider the within-ensemble distance distribution as a function of the type of graph and model parameters. While our focus is on the means of the distance distributions, we also include the standard deviations in each figure. Ultimately, we report the within-ensemble graph distances for 20 different graph distances from the software package, \texttt{netrd} \footnote{\url{https://github.com/netsiphd/netrd/}. Note: this software package includes several more distances that were not included in these analyses, and as it is an open-source project, we anticipate that it will be updated with new distance measures as they continue to be developed.}. To our knowledge, this is the largest systematic comparison of graph distances to date.

\section{Methods}\label{sec:meth}

\subsection{Ensembles}\label{sec:meth_ensembles}
We study the behavior of $(d, \Psi, \mathcal{M})$ for sets of graphs sampled from $M_{\vec{\alpha}}$ under a variety of parameterizations. There are many graph ensembles that one could use to compute within-ensemble graph distances, and we begin by focusing on two broad classes: ensembles that produce graphs with homogeneous degree distributions and those that produce graphs with heterogeneous degree distributions. In total, we study the within-ensemble graph distance for five different ensembles.

\subsubsection{Erd\H{o}s-Rényi random graphs}\label{sec:meth_densegnp}

Graphs sampled from the Erd\texorpdfstring{\H{o}}{Lg}s-Rényi model ($ER$), also known as $G_{(n,p)}$, have (undirected) edges among $n$ nodes, with each pair being connected with probability $p$ \cite{Erdos1959, Bollobas1980AGraphs}. This model is commonly used as a benchmark or a null model to compare with observed properties of real-world network data from nature and society. In our case, it allows us to explore the behavior of graph distance measures on dense and homogeneous graphs without any structure. In fact, this model maximizes entropy subject to a global constraint on expected edge density, $p$.

One well-studied construction of this ensemble is when $p=\frac{\langle k \rangle}{n}$, in which $n$ nodes are connected uniformly at random such that nodes in the resulting graph have an average degree of $\langle k \rangle$. This ensemble is particularly useful for identifying which graph distance measures are able to capture key structural transitions that happen as the average degree increases. For convenience, we will refer to this ensemble as $G_{(n,\langle k \rangle)}$.

\subsubsection{Random geometric graphs}\label{sec:meth_densergg}

We work with random geometric graphs of $n$ nodes and edge density $p$, generated by sprinkling $n$ coordinates uniformly into a one-dimensional ring of circumference $1$, and connecting all pairs of nodes whose coordinate distance (arc length) is less than or equal to $\frac{p}{2}$. Compared to $G_{(n,p)}$, this model produces graphs that have a high average local clustering coefficient, which is a property commonly found in real network data. Note that setting the connection distance to $\frac{p}{2}$ means that $p$ parameterizes the edge density exactly as in $G_{(n,p)}$ \cite{Dall2002, Penrose2003}.

\subsubsection{Watts-Strogatz graphs}\label{sec:meth_wattsstrogatz}

Watts-Strogatz ($WS$) graphs allow us to study the effects that random, long-range connections have on otherwise large-world regular lattices. A $WS$ graph is initialized as a one-dimensional regular ring-lattice, parameterized by the number of nodes $n$ and the even-integer degree of every node $\langle k \rangle$ (each node connects to the $\frac{\langle k\rangle}{2}$ closest other nodes on either side). Each edge in the network is then randomly rewired with probability $p_r$, which generates graphs with both relatively high average clustering and relatively short average path lengths for a wide range of $p_r\in(0,1)$ \cite{Watts1998}.

\subsubsection{(Soft) Configuration model with power-law degree distribution}\label{sec:meth_power-lawCM} 

We generate \textit{expected} degree sequences from distributions with power-law tails with a mean of $\langle k \rangle$. We construct an instance of a ``soft'' configuration model, the maximum entropy network ensemble with a given sequence of expected degrees, by connecting node-pairs with probabilities determined via the method of Lagrange multipliers \cite{Park2004, Garlaschelli2008, Cimini2019}. Through this method, we are able to construct networks with a tunable degree exponent, $\gamma$. The degree exponents that we test range from those that skew the distribution heavily, resulting in a highly heterogeneous ultra-small-world network ($\gamma\in(2,3)$), to those that generate more homogeneous networks ($\gamma > 3$). In contrast to the homogeneous ensembles we tested---all of which have homogeneous degree distributions---the requirement of heterogeneity in these graphs constrains the possible edge densities to be vanishingly small. Otherwise, in the high-edge density regime, degrees cannot fluctuate to appreciably larger-than-average values, and we have a natural degree scale imposed by the network size.

\subsubsection{Nonlinear preferential attachment}\label{sec:meth_prefattach}

The final ensemble of networks included here are grown under a degree-based nonlinear preferential attachment mechanism \cite{Albert1999, Albert2002, Krapivsky2000}. A network of $n$ nodes is grown as follows: each new node is added to the network sequentially, connecting its $m$ edges to nodes already in the network $v_i \in V$ with probability $\Pi_i = \frac{k_i^\alpha}{\sum_j k_j^\alpha}$, where $k_i$ is the degree of node $v_i$ and $\alpha$ modulates the probability that a given node already in the network will collect new edges. When $\alpha=1$, this model generates networks with a power-law degree distribution (with degree exponent $\gamma=3$), and a condensation regime emerges as $n \rightarrow \infty$ when $\alpha>2$, producing a star network with $O(n)$ nodes all connected to a main hub node \cite{Krapivsky2000}.

\subsection{Graph distance measures}\label{sec:meth_dists}
The study of network similarity and graph distance has yielded many approaches for comparing two graphs \cite{Donnat2018}. Typically, these methods involve comparing simple descriptors based on either aggregate statistical properties of two graphs---such as their degree or average path length distributions \cite{Bagrow2019}---or intrinsic spectral properties of the two graphs, such as the eigenvalues of their adjacency matrices, or of other matrix representations \cite{Jurman2011}. The description distances also tend to fall in two broad categories: either classic definitions of norms or distances based on statistical divergence. While different approaches are better suited for capturing differences between certain types of graphs, they obviously are expected to share several properties.

The simplest graph distances aggregate element-wise comparisons between the adjacency matrices of two graphs \cite{Golub2013, Jaccard1901, Hamming1950, gao2010survey}, and extensions thereof \cite{Wallis2001GraphUnion}; these methods depend explicitly on the node labeling scheme (and hence are not invariant under graph isomorphism \cite{Chowdhury2017DistancesInvariants}), which may limit their utility when comparing graphs with unknown labels (e.g. graphs sampled from random graph ensembles, as we do here). Several measures collect empirical distributions \cite{Carpi2011} or a ``signature'' vector \cite{berlingerio2012netsimile} from each graph and take the distance between them (using the Jensen-Shannon divergence, Canberra distance, earth mover's distance, etc. \footnote{From our preliminary analyses, the particular choice of metric can dramatically change the distance values, though we do not report this here. For an extensive description of distance metrics in general, see \cite{deza2009encyclopedia, EmmertStreibEtAl2016}.}), which, among other things, facilitates comparison of differently sized graphs \cite{Bagrow2019, Meila2007}. Another family of approaches compare spectral properties of certain matrices characterized by the graphs \cite{Jurman2015}, such as the non-backtracking matrix \cite{Torres2019, Mellor2019} or Laplacian matrix \cite{Jurman2011}. The relevant spectral properties associated with these distances are invariant under graph isomorphism \cite{VanSteen2010, Chowdhury2017DistancesInvariants}. Some graph distances have been shown to be metrics (i.e., they satisfy properties such as triangle inequality, etc.) \cite{Bento2019}, whereas others have not. These are not exhaustive descriptions of every graph distance in use today, but they represent coarse similarities between the various methods. We summarize the 20 graph distances we consider in Table \ref{table:dists} and more extensively define them in Supplemental Information (SI) \ref{sec:SI_distances}.

\begin{table}[t!]
\centering
\begin{tabular}{clcc}
 & \textbf{Graph distance} & \textbf{Label} & \\
\hline
1 & Jaccard \cite{Jaccard1901} & \texttt{JAC} & \\ 
2 & Hamming \cite{Hamming1950} & \texttt{HAM} & \\
3 & Hamming-Ipsen-Mikhailov \cite{Jurman2015} & \texttt{HIM} &  \\ 
4 & Frobenius \cite{Golub2013} & \texttt{FRO} &  \\ 
5 & Polynomial dissimilarity \cite{Donnat2018} & \texttt{POD} &  \\ 
6 & Degree JSD \cite{Carpi2011} & \texttt{DJS} &  \\
7 & Portrait divergence \cite{Bagrow2019} & \texttt{POR} &  \\
8 & Quantum spectral JSD \cite{DeDomenico2016SpectralComparison} & \texttt{QJS} &  \\ 
9 & Communicability sequence \cite{Chen2018}  & \texttt{CSE} & \\
10 & Graph diffusion distance \cite{Hammond2013} & \texttt{GDD} & \\
11 & Resistance-perturbation \cite{Monnig2018} & \texttt{REP} &  \\ 
12 & NetLSD \cite{Tsitsulin2018} & \texttt{LSD} & \\ 
13 & Lap. spectrum; Gauss. kernel, JSD \cite{Jurman2011}& \texttt{LGJ}  &   \\ 
14 & Lap. spectrum; Loren. kernel, Euc. \cite{Jurman2011}& \texttt{LLE} &  \\ 
15 & Ipsen-Mikhailov \cite{Ipsen2002}  & \texttt{IPM} &\\ 
16 & Non-backtracking eigenvalue \cite{Torres2019} & \texttt{NBD} &  \\ 
17 & Distributional Non-backtracking \cite{Mellor2019} & \texttt{DNB} &  \\ 
18 & D-measure distance \cite{Schieber2017}  & \texttt{DMD} & \\
19 & DeltaCon \cite{koutra2016deltacon} & \texttt{DCN} &  \\ 
20 & NetSimile \cite{berlingerio2012netsimile} & \texttt{NES} &  \\ 
\end{tabular}
\caption{\textbf{Graph distances.} Distance measures used to systematically compare graphs in this work, as well as their abbreviated labels, and their source. \textit{Abbreviations}: Lap. = Laplacian, Gauss. = Gaussian, Loren. = Lorenzian, JSD = Jensen-Shannon divergence, Euc. = Euclidian distance.}
\label{table:dists}
\end{table}

\subsection{Description of experiments}\label{sec:experiments}

\begin{table}[t!]
\centering
\begin{tabular}{lll}
 \textbf{Ensemble} & \textbf{Fixed parameter(s)} & \textbf{Key parameter} \\
\hline
$G_{(n,p)}$ & $n = 500$ & $p \in \{0.02, 0.06, ..., 0.98\}$\\
$RGG$ & $n = 500$ & $p \in \{0.02, 0.06, ..., 0.98\}$\\
$G_{(n,\langle k \rangle)}$ & $n = 500$ & $\langle k \rangle \in \{10^{-4},...,n\}$\\
$WS$ & $n = 500, \langle k \rangle=8$ & $p_r \in \{10^{-4},...,10^{0}\}$\\
$SCM$ & $n = 1000, \langle k \rangle=12$ & $\gamma \in \{2.01, 2.06, ... 6.01\}$\\
$PA$ & $n = 500, \langle k \rangle=4$ & $\alpha \in \{-5, -4.95, ..., 5\}$\\ \hline
\end{tabular}
\caption{\textbf{Experiment parameterization.} Here we report the ensembles that were used in these experiments, as well as their parameterizations. For $G_{(n,\langle k \rangle)}$ and $WS$ key parameters, we span 100 values, spaced logarithmically, between the values above. \textit{Parameter labels}: $n$ = network size, $p$ = density, $\langle k \rangle$ = average degree, $p_r$ = probability that a random edge is randomly rewired, $\gamma$ = power-law degree exponent, $\alpha$ = preferential attachment kernel. Note: In SI \ref{sec:wegd_n}, we show how the within-ensemble graph distance changes as $n$ increases.}
\label{table:ens_params}
\end{table}
See Table \ref{table:ens_params} for the full parameterization of these sampled graphs. In each experiment, we generate $N = 10^3$ pairs of graphs for every combination of parameters. With these sampled random graphs, we measure the distance between pairs from the same parameterization of the same model, $M_{\vec{\alpha}}$, and report statistical properties of the resulting vectors of distances. In other words, our experiments consist of calculating mean within-ensemble graph distances,
\begin{equation}
    \langle D \rangle = \sum_{G,G'\in\mathcal{G}}D(G, G')\mathbb{P}_{\vec\alpha}(G)\mathbb{P}_{\vec\alpha}(G'),
\end{equation}
where $\mathbb{P}_{M, \vec{\alpha}}:\mathcal{G}\rightarrow[0,1]$ (or $\mathbb{P}_{\vec{\alpha}}$ when its meaning is unambiguous) is the graph probability distribution for model $M_{\vec{\alpha}}$. This is estimated by sampling $N\gg 1$ graph-pairs $\{(G_i,G_i')\}_{i=1}^N$ and computing
\begin{equation}
    \langle D \rangle \approx \frac{1}{N}\sum_{i=1}^N D(G_i, G_i')\;.
\end{equation}

We then study the behavior of $\langle D \rangle$ for various $M_{\vec{\alpha}}$. The error on the mean within-ensemble graph distance is estimated from the following standard error of the mean $\sigma_{\langle D \rangle} \approx \frac{\sigma_D}{\sqrt{N}}$, where $\sigma_D$ is the standard deviation on the within-ensemble graph distance $D$, estimated by sampling as well.
For all experiments, we used $N = 10^3$ pairs of graphs, which is sufficient in general as can be seen from the small standard error relative to the mean in all figures. In each plot, we also include the standard deviations $\sigma_D$ of the within-ensemble graph distances, and we highlight when the standard deviation offers particularly notable insights into the behavior of certain distances.

Lastly, there are several distances that assume alignment in the node labels of $G$ and $G'$. Because we are sampling from random graph ensembles, the networks we study here are not node-aligned, and as such, care should be taken when interpreting the output of these graph distances. For every description of graph distances in SI \ref{sec:SI_distances}, we note if node alignment is assumed. 

\section{Results}\label{sec:results}

In the following sections, we broadly describe the behavior of the mean within-ensemble graph distance (in general denoted $\langle D \rangle$) for the distance measures tested. The general structure of this section is motivated by critical properties of the ensembles studied here. We highlight features of the within-ensemble graph distance for two broad characterizations of networks: homogeneous and heterogeneous graph ensembles, focusing on specific ensembles within each category.

All of the main results from the experiments described below are summarized in Table \ref{table:summary}, which practitioners may find especially useful when considering which tools to use for comparing networks with particular structures. When relevant, we highlight certain distance measures to emphasize interesting within-ensemble graph distance behaviors.

\begin{table*}[ht]
\resizebox{\textwidth}{!}{
\centering 
    \begin{tabular}{l | l | c c c c c c c c c c c c c c c c c c c c} 
        \hline\hline
        \textbf{Model} & \textbf{Property} & \texttt{JAC} & \texttt{HAM} & \texttt{HIM} & \texttt{FRO} & \texttt{POD} & \texttt{DJS} & \texttt{POR} & \texttt{QJS} & \texttt{CSE} & \texttt{GDD} & \texttt{REP} & \texttt{LSD} & \texttt{LGJ} & \texttt{LLE} & \texttt{IPM} & \texttt{NBD} & \texttt{DNB} & \texttt{DMD} & \texttt{DCN} & \texttt{NES}  \\  
        \hline
        $G_{(n,p)}$  & Complement symmetry  &  & $\checkmark$  &  $\checkmark$ & $\checkmark$  & $\checkmark$ & $\checkmark$ &  &  &  &  &  &  &  &  &  &  &  &   \\ [0.5ex]
        $G_{(n,p)}$  & Derivative with network size, $n$ & 0 & 0 & 0 & $+$ & $-$ & $-$ & $-$ & $\sim$ & $-$ & $\sim$ & $-$ & $-$ & $-$ & $-$ & $-$ & $+$ & $-$ & $-$ & $+$ &  $\sim$ \\ [0.5ex]
        \hline
        $RGG$  & Maximum: $p\approx\frac{1}{2}$ &  & $\checkmark$ & $\checkmark$ & $\checkmark$ & $\checkmark$ & $\checkmark$ & $\checkmark$ &  &  &  &  &  &  &  &  $\checkmark$ &  &  &  &   \\ [0.5ex] 
        \hline
        $G_{(n,\langle k \rangle)}$  & Detects the giant 1-core &  &  &  &  &  & & $\checkmark^*$ & & $\checkmark^*$  & $\checkmark$ &  &$\checkmark$   &$\checkmark^*$  &$\checkmark$  &$\checkmark$  & $\checkmark^*$ & $\checkmark$   & $\checkmark^*$ &  &   \\ [0.5ex] 
        $G_{(n,\langle k \rangle)}$ & Detects the giant 2-core &  &  &  &  &  &  &  $\checkmark^*$ &  & $\checkmark^*$ &  &  &  &  &  &  &  &  & $\checkmark^*$ &  & $\checkmark^*$  \\ [0.5ex] 
        $G_{(n,\langle k \rangle)}$  & Derivative with network size, $n$ & $0$ & $-$ & $-$ & $+$ & $-$ & $-$ & $\sim$ & $0$ & $-$ & $+$ & $+$ & $+$ & $-$ & $-$ & $-$ & $\sim$ & $-$ & $-$ & $+$ & $-$ \\ [0.5ex] 
        \hline
        $WS$  & Small-world $>$ random &  &  &  &  &  &  & $\checkmark$ &  & $\checkmark$ &  & $\checkmark$ & $\checkmark$ & $\checkmark$ & $\checkmark$ & $\checkmark$ &  &  & $\checkmark$ & &  \\ [0.5ex] 
        $WS$  & Path length sensitivity &  &  &  &  &  & $\checkmark^*$ & $\checkmark$ &  &  & $\checkmark$ & $\checkmark$ & & $\checkmark^*$ & & & $\checkmark$ & $\checkmark$ & $\checkmark$ & $\checkmark^*$ & $\checkmark$   \\ [0.5ex]
        $WS$  & Clustering sensitivity  &  &  &  &  &  &  &  &  & $\checkmark$ & $\checkmark$ &  & $\checkmark$ & $\checkmark$ & $\checkmark$ & $\checkmark$ &  & $\checkmark^*$ &  & & $\checkmark^*$  \\ [0.5ex] 
        \hline
        $SCM$  & Maximum: $2<\gamma<3$  & $\checkmark$ & $\checkmark$ & $\checkmark$ & $\checkmark$ & $\checkmark$ & $\checkmark$ & $\checkmark$ & $\checkmark$ & $\checkmark$ & $\checkmark$ & $\checkmark$ & $\checkmark$ & $\checkmark$ & $\checkmark$ & $\checkmark$ & $\checkmark$ & $\checkmark$ & $\checkmark$ &  & $\checkmark$\\ [0.5ex] 
        $SCM$  & Monotonic decay as $\gamma$ grows & & $\checkmark^\dagger$ & $\checkmark$ & $\checkmark^\dagger$ & $\checkmark^\dagger$ & $\checkmark$ & $\checkmark^\dagger$ & $\checkmark^\dagger$ & & $\checkmark$ &  & $\checkmark$ & $\checkmark$ & $\checkmark$ & $\checkmark$ & $\checkmark^\dagger$ & $\checkmark$ &  $\checkmark$  & & $\checkmark^\dagger$\\ [0.5ex] 
        \hline 
        $PA$  & Heterogeneous $>$ homogeneous &  &  &  &  & &  &  &  & $\checkmark$ & $\checkmark$ &  &  & &  &  &  &  &  & &  \\ [0.5ex] 
        $PA$  & Maximum: $\alpha\approx0$ (uniform) & $\checkmark$ & $\checkmark$ &  & $\checkmark$ & $\checkmark$ &  &  & &  & &  &  & &  &  &  &  & & &  \\ [0.5ex] 
        $PA$  & Maximum: $\alpha\approx1$ (linear) &  &  &  &  &  &  & & $\checkmark$ &  &  & $\checkmark$ & & &  & &  &  &  & &  \\ [0.5ex]
        $PA$  & Maximum: $1 < \alpha \leq 2$ &  &  & $\checkmark$ &  &  &$\checkmark$ & $\checkmark$ & & $\checkmark$ &  &  &  $\checkmark$ & $\checkmark$ & $\checkmark$ & $\checkmark$ &  $\checkmark$ & $\checkmark$ & $\checkmark$ & & $\checkmark$ \\ [0.5ex] 
        \hline
    \end{tabular}}
    \parbox{\linewidth}{\raggedright
        \hspace{0.5cm}\scriptsize $\checkmark$ = captures a given property through a global maximum/minimum in its within-ensemble graph distance curve.\\
        \hspace{0.5cm}\scriptsize $\sim$ = non-monotonic relationship between network size and within-ensemble graph distance.\\
        \hspace{0.5cm}\scriptsize $\checkmark^*$ = potentially captures a given property (via local maxima in the mean or standard deviation, change in slope, etc.).\\
        \hspace{0.5cm}\scriptsize $\checkmark^\dagger$ = monotonic decay beyond a very small value of $\gamma$ ($\gamma\approx 2$) where there is an apparent maximum (for $SCM$).
    }
    \caption{\textbf{Summary of key within-ensemble graph distance properties for different ensembles.} Each of the ensembles included in this work has characteristic properties that a within-ensemble graph distance may be able capture. Here we consolidate these various properties into a single table that classifies whether each distance has a given property. Models considered are dense Erd\H{o}s-R\'enyi graphs ($G_{(n,p)}$), random geometric graphs ($RGG$), sparse Erd\H{o}s-R\'enyi graphs ($G_{(n,\langle k\rangle)}$), the Watts-Strogatz model ($WS$), soft configuration model with power-law degree distribution ($SCM$) and general preferential attachment with kernel $\alpha$ ($PA$). \textit{Clarifications}: In the $WS$ model, we look at three properties: 1) the mean within-ensemble graph distance is larger for intermediate ``small-world'' values of $p_r$ than it is when $p_r=1$; 2) the within-ensemble graph distance is sensitive to values of $p_r$ where the magnitude slope of the $L_p/L_0$ curve is largest (``path length sensitivity'' above); 3) the within-ensemble graph distance is sensitive to values of $p_r$ where the magnitude slope of the $C_p/C_0$ curve is largest (``clustering sensitivity'' above). In the $PA$ model, we look at whether high, positive values of $\alpha$ produce greater mean within-ensemble graph distances than lower, negative values of $\alpha$, and at where the maximum within-ensemble distance occurs.}
    \label{table:summary}
\end{table*}

\begin{figure*}[t!]
    \centering
    \includegraphics[width=1.0\textwidth]{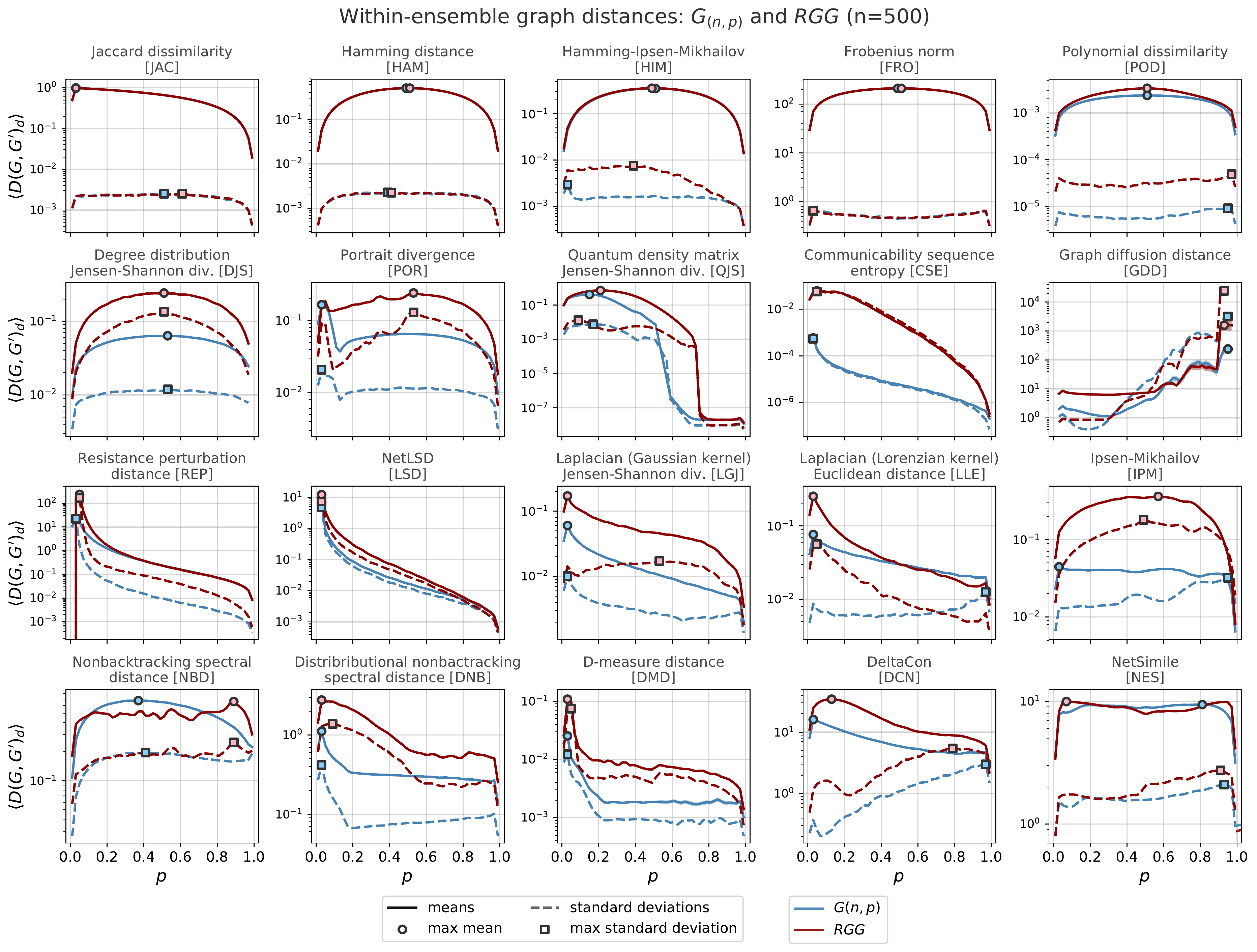}
    \caption{\textbf{Mean and standard deviations of the within-ensemble distances for $G_{(n,p)}$ and $RGG$.} By repeatedly measuring the distance between pairs of $G_{(n,p)}$ and $RGG$ networks of the same size and density, we begin to see characteristic behavior in both the graph ensembles as well as the graph distance measures themselves. In each subplot, the mean within-ensemble graph distance is plotted as a solid line with a shaded region around for the standard error ($\langle D \rangle \pm \sigma_{\langle D \rangle}$; note that in most subplots  above, the standard error is too small to see), while the dashed lines are the standard deviations.}
    \label{fig:gnp_rgg_wegd}
\end{figure*}

\begin{figure*}[t!]
    \centering
    \includegraphics[width=1.0\textwidth]{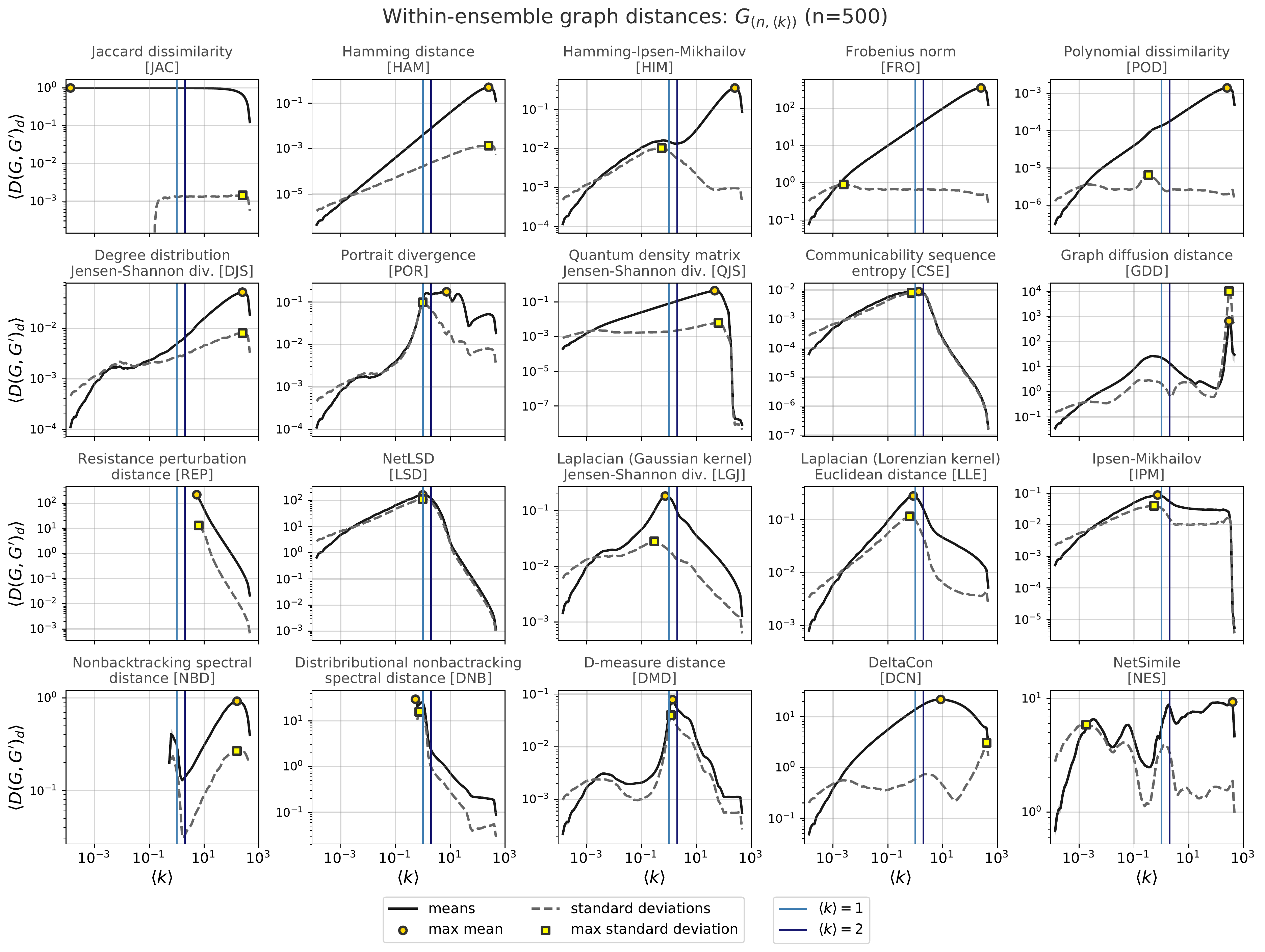}
    \caption{\textbf{Mean and standard deviations of the within-ensemble distances for $G_{(n,\langle k \rangle)}$ networks.} Here, we generate pairs of $ER$ networks with a given average degree, $\langle k \rangle$, and measure the distance between them with each distance measure. In each subplot, we highlight $\langle k \rangle = 1$ and $\langle k \rangle = 2$. In each subplot, the mean within-ensemble graph distance is plotted as a solid line with a shaded region around for the standard error ($\langle D \rangle \pm \sigma_{\langle D \rangle} $; note that in most subplots  above, the standard error is too small to see), while the dashed lines are the standard deviations.} 
    \label{fig:gnk_wegd}
\end{figure*}

\begin{figure*}[t!]
    \centering
    \includegraphics[width=1.0\textwidth]{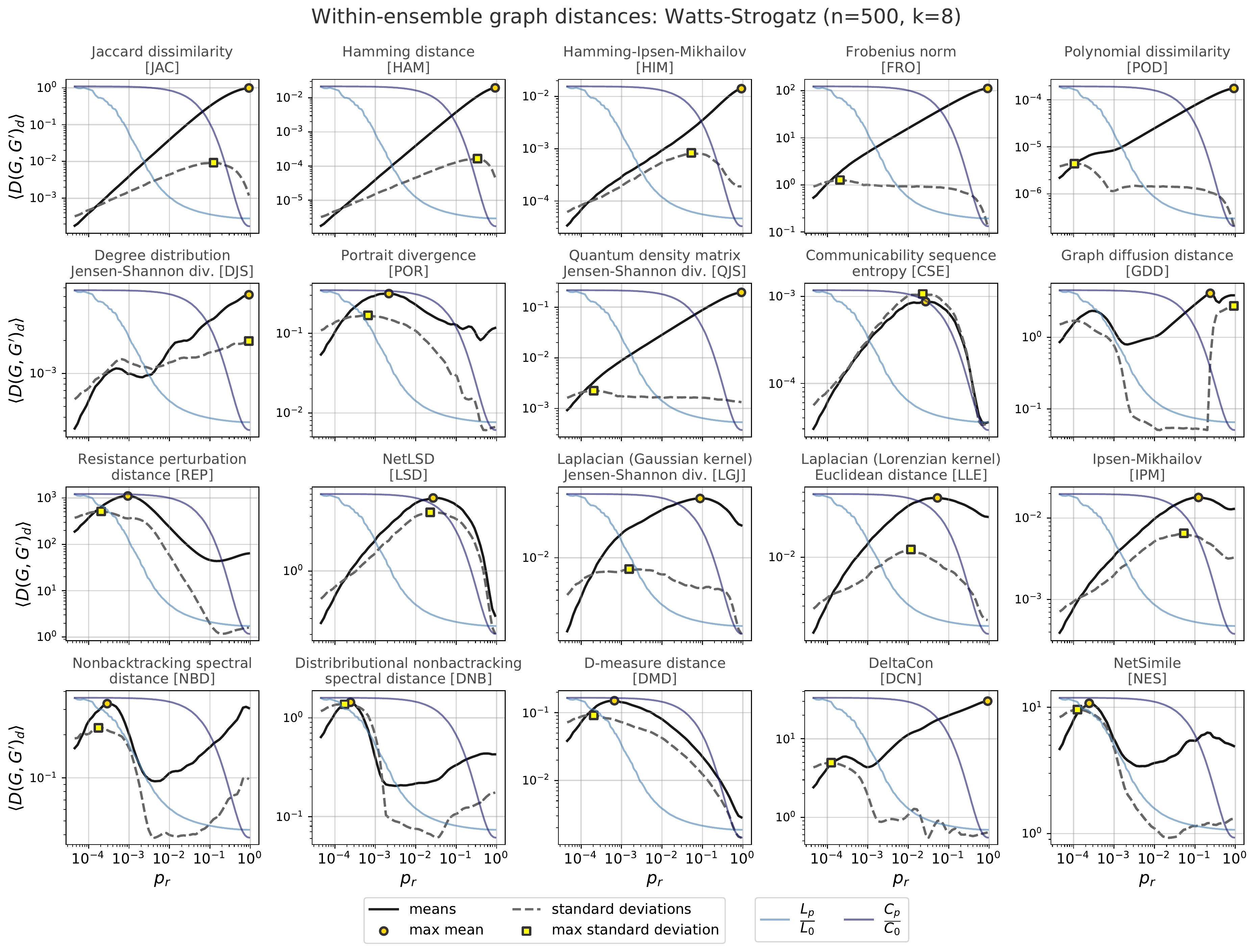}
    \caption{\textbf{Mean and standard deviations of the within-ensemble distances for Watts-Strogatz networks.} Here, we generate pairs of Watts-Strogatz networks with a fixed size and average degree but a variable probability of rewiring random edges, $p_r$. In each subplot we also plot the clustering and path length curves as in the original Watts-Strogatz paper \cite{Watts1998} to accentuate the ``small-world'' regime with high clustering and low path lengths. The mean within-ensemble graph distance is plotted as a solid line with a shaded region around for the standard error ($\langle D \rangle \pm \sigma_{\langle D \rangle} $; note that in most subplots  above, the standard error is too small to see), while the dashed lines are the standard deviations.}
    \label{fig:ws_wegd}
\end{figure*}

\subsection{Results for homogeneous graph ensembles}\label{sec:results_homogeneous}

\subsubsection{Dense graph ensembles}\label{sec:results_dense}

Here, we present our results for the two models that produce homogeneous and dense graphs. 

The $G_{(n,p)}$ model possesses three notable features that we might expect graph distance measures to recover. Note that while we might expect graph distances to recover these features, we are not asserting that every graph distance measure \textit{should} capture these properties.

\begin{enumerate}

\item The size of the ensembles shrink to a single isomorphic class in the limits $p \rightarrow 0$ and $p \rightarrow 1$, corresponding respectively to an empty and complete graph of size $n$. In both limits, we might therefore expect $\langle D(M_{n,p}) \rangle$ to go to zero for any method that considers unlabelled graphs. 

\item The $G_{(n,p)}$ model creates ensembles of graphs and graph complements symmetric under the change of variable $p'=1-p$. By definition, every graph $G$ has a complement $\bar{G}$ such that every edge that does (or does not) exist in $G$ does not (or does) exist in $\bar{G}$. Therefore, for every graph in $G_{(n,p)}$, one can expect to find its complement occurring with the same probability in $G_{(n,1-p)}$. We might expect $\langle D(M_{n,p}) \rangle = \langle D(M_{n,1-p}) \rangle$ if graph distances can capture this symmetry.

\item A density of $p=\frac{1}{2}$ produces the $G_{(n,p)}$ ensemble with maximal entropy (all graph configurations have an equal probability). As a result, we might also expect $\langle D(M_{n,p}) \rangle$ to have a global maximum at $p=\frac{1}{2}$.

\end{enumerate}

The $RGG$ model shares features 1 and 3 with the $G_{(n,p)}$ model, but not feature 2. Moreover, the most significant differences between the two models is that edges are not independent in the $RGG$ model. Correlations between edges lead to local structure (i.e., higher-order structures like triangles) and to correlations in the joint-degree distribution. We therefore do not expect distance measures focused on the degree distribution to produce exactly the same mean within-ensemble distance curve in $RGG$ as in $G_{(n,p)}$. Conversely, any distance measure that does produce the exact same within-ensemble distance curve for $RGG$ and $G_{(n,p)}$ either fails to account for these correlations, or the effect of these correlations is negligible on the overall distance between two graphs drawn from the ensemble. This is the case for \texttt{HAM}, \texttt{HIM} and \texttt{FRO}.

Our result for homogeneous graph ensembles are shown in Figure \ref{fig:gnp_rgg_wegd}. Only 5 out of 20 graph distances capture all features discussed above, namely: \texttt{HAM}, \texttt{HIM}, \texttt{FRO}, \texttt{POD}, \texttt{DJS}. Notably, these are some of the simplest methods considered. In fact, these include two in which theoretical predictions for $ER$ graphs precisely match the observed results for both $ER$ graphs and $RGG$s, despite no consideration of $RGG$s having been included in such calculations. In one such case (\texttt{FRO}), $ER$ graphs and $RGG$s behave identically, yet there is also an $n$-dependence (See SI Figure \ref{fig:n_wegd}).

\subsubsection{Sparse graph ensembles}\label{sec:results_sparse}

While the previous section highlighted dense $RGG$ and $ER$ networks, we now turn to the within-ensemble graph distance of \textit{sparse} homogeneous graphs sampled from $G_{(n,p)}$, such that $p=\frac{\langle k \rangle}{n}$. In the case of sparse graphs, the edge density decays to zero in the $n\rightarrow\infty$ limit as the mean degree $\langle k \rangle$ remains fixed. We found it important to cast this distinction between dense $G_{(n,p)}$ because of critical transitions that take place as $\langle k \rangle$ increases. As network scientists, these early transition points in sparse networks are foundational, with implications for a number of network phenomena (i.e. the occurrence of outbreaks in disease models \cite{molloy_critical_1995}, etc.).

In fact, the presence of such critical transitions in random graph models underscores the utility of this approach for studying graph distance measures. That is, a sudden change in the within-ensemble graph distance signals abrupt changes in the probability distribution over the set of graphs in the ensemble (i.e., the emergence of novel graph structures that are markedly different from the greater population of graphs in an ensemble). This may show up as a local or global maximum within-ensemble graph distance near parameter values for which this transition occurs. Conversely, if a sudden decrease in within-ensemble graph distance is observed, then there may be a sudden disappearance or reduction in largely dissimilar graphs in the ensemble.

In the case of $G_{(n,p)}$ where $p=\frac{\langle k \rangle}{n}$, which we will refer to with the shorthand, $G_{(n,\langle k \rangle)}$, the following critical transitions emerge:
\begin{enumerate}
    \setcounter{enumi}{3}
    \item At $\langle k \rangle=1$, we see the emergence of a giant component in $ER$ networks (likewise, a 2-core emerges at $\langle k \rangle=2$). We might expect, for example, a within-$G_{(n,\langle k \rangle)}$ graph distance to have a local maximum at such values. 
\end{enumerate}

Ultimately, we observe that distance measures that are fundamentally associated with flow-based properties of the network (i.e., if a distance measure is based on a graph's Laplacian matrix, communicability, or other properties important to diffusion, such as path-length distributions, etc.) are the ones most sensitive for picking up on this property (Figure \ref{fig:gnk_wegd}) \footnote{Note that the two distance measures based on the non-backtracking matrix (\texttt{NBD} \& \texttt{DNB}) are undefined in graphs without a 2-core, restricting their range in Figure \ref{fig:gnk_wegd}.}.

What Figure \ref{fig:gnk_wegd} highlights, which the dense ensembles in Figure \ref{fig:gnp_rgg_wegd} could not, is the rich and varied behavior characteristic of sparse graphs. For example, the distance measures with maxima at $p=\frac{1}{2}$ (\texttt{HAM}, \texttt{HIM}, \texttt{FRO}, \texttt{POD}, \texttt{DJS}, etc.) are still seen in Figure \ref{fig:gnk_wegd}, but the emphasis is instead on the degree as opposed to the edge density; given that most real-world networks are sparse \cite{DelGenio2011}, this view of the same parameter is especially informative.

Importantly, while the qualitative behaviors discussed here are general features of the models and distances, the quantitative value of the average within-ensemble graph distance also depends on network size. There are no specific structural transitions to discuss around this dependency, but it can be an important problem when comparing networks of different sizes without a good understanding of how network distances might behave. Interested readers can find our results in SI \ref{sec:wegd_n} where we use $G_{(n,\langle k \rangle)}$ to vary network size while keeping all other features fixed.

\subsubsection{Small-world graphs}\label{sec:results_regular}
The final homogeneous graph ensemble studied here is the Watts-Strogatz model. This model generates networks that are initialized as lattice networks, and edges are randomly rewired with probability, $p_r$. At certain values of $p_r$, we see two key phenomena occur:
\begin{enumerate}
    \setcounter{enumi}{4}
    \item ``Entry'' into the small-world regime: Even as the edges in the network are minimally rewired, the average path length quickly decreases relative to its initial (longer) value. This is highlighted by the blue curve in Figure \ref{fig:ws_wegd}, corresponding to $\frac{L_p}{L_0}$, where $L_0$ is the average path length before any edges have been rewired. For the parameterizations used in this study, the largest (negative) slope of this curve is at $p_r \approx 2 \times 10^{-3}$. We might expect a within-ensemble graph distance to be sensitive to this or nearby values of $p_r$, as this region corresponds to changes in the graphs' common structural features.
    \item ``Exit'' from the small-world regime: After enough edges have been rewired, the network loses whatever clustering it had from originally being a lattice, reducing to approximately the clustering of an $ER$ graph. This is highlighted by the violet curve in Figure \ref{fig:ws_wegd}, corresponding to $\frac{C_p}{C_0}$, where $C_0$ is the average clustering before any edges have been rewired. For the parameterizations used in this study, the largest (negative) slope of this curve is at $p_r \approx 3 \times 10^{-1}$. Again, we might expect a within-ensemble graph distance to be sensitive to this large decrease in clustering.
\end{enumerate}

Together, the above features characterize Watts-Strogatz networks. Importantly, we are interested in whether a distance measure is \textit{sensitive} to these ``entry'' and ``exit'' values of $p_r$; sensitive here is deliberately broadly defined. For instance, as in the case of \texttt{CSE}, we observe a reduction in within-ensemble graph distance at a rate that almost exactly resembles the rate at which $\frac{C_p}{C_0}$ decays. Alternatively, a distance measure can be sensitive to these critical points by having a local maximum at or around the critical point. In the case of \texttt{POR}, we see that the within-ensemble graph distance is maximized at approximately the same point as the largest (negative) slope of the $\frac{L_p}{L_0}$ curve.

Here, \textit{insensitivity} to these critical points is also an informative property to highlight in a distance measure. As one example, \texttt{HAM} appears to be otherwise unaffected by the ``exit'' from the small-world regime, with distances increasing steadily despite the model generating networks with dramatic structural differences.

Lastly, we ask whether the within-ensemble graph distance of random networks (i.e., when $p_r \rightarrow 1$) is greater than that of small-world networks; this is indicated by a within-ensemble graph distance curve that is higher at $p_r=1$ than those between $10^{-3} < p_r < 10^{-1}$ in Figure \ref{fig:ws_wegd}. This property holds for distance measures that depend on node labeling (e.g. \texttt{JAC}, \texttt{HAM}, \texttt{HIM}, \texttt{FRO}, \texttt{POD}, etc.) but also for \texttt{DJS}---which is intuitive, since more noise increases the variance of the degree distribution---as well as a few puzzling distances: \texttt{QJS}, \texttt{DCN}, and the two based on the non-backtracking matrix, \texttt{NBD} and \texttt{DNB}. 

\begin{figure*}[t!]
    \centering
    \includegraphics[width=1.0\textwidth]{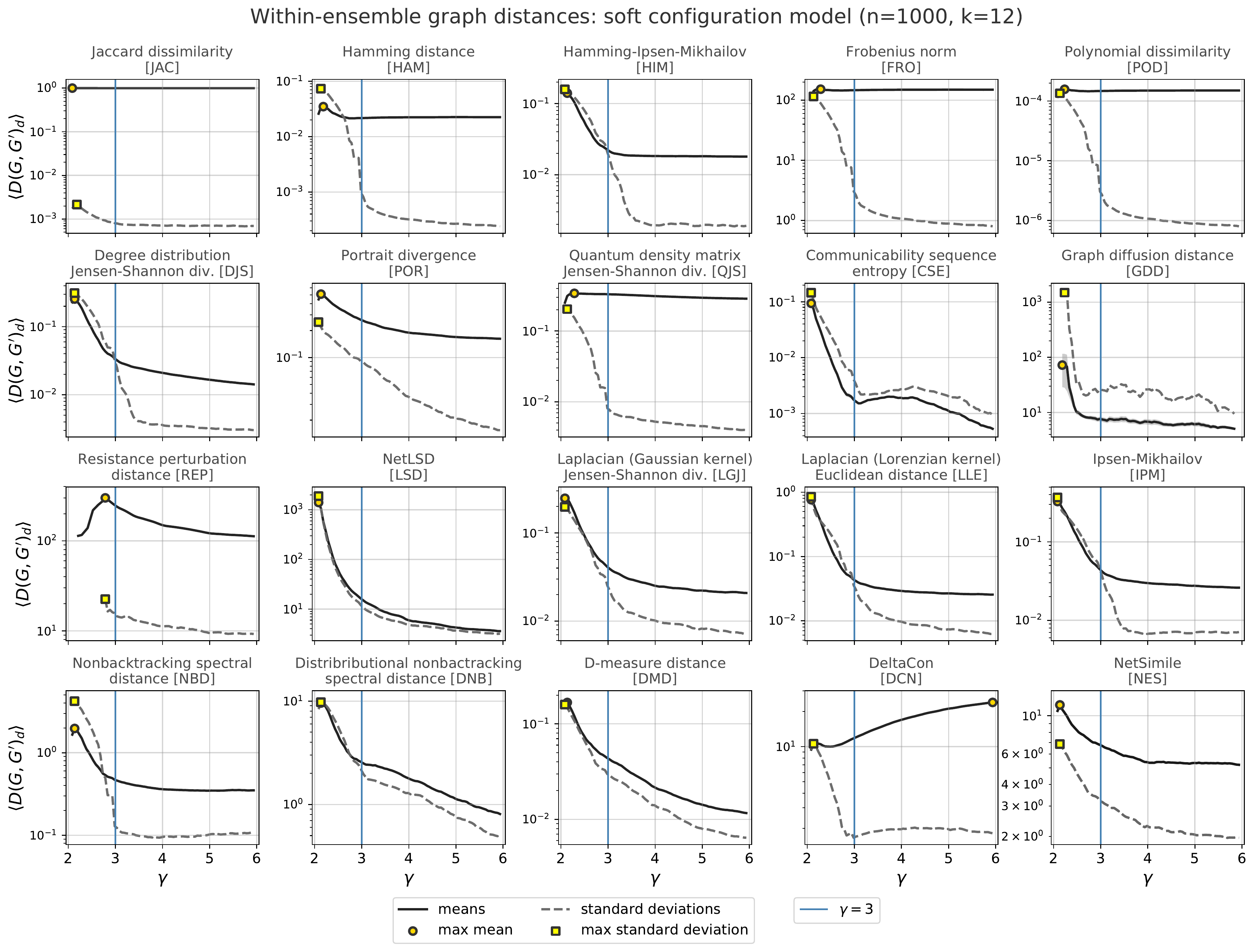}
    \caption{\textbf{Mean and standard deviations of the within-ensemble distances for soft configuration model networks with varying degree exponent.} Here, we generate pairs of networks from a (soft) configuration model, varying the degree exponent, $\gamma$, while keeping $\langle k \rangle$ constant ($n=1000$). In each subplot we highlight $\gamma=3$. The mean within-ensemble graph distance is plotted as a solid line with a shaded region around for the standard error ($\langle D \rangle \pm \sigma_{\langle D \rangle} $; note that in most subplots  above, the standard error is too small to see), while the dashed lines are the standard deviations.}
    \label{fig:scm_wegd}
\end{figure*}

\begin{figure*}[t!]
    \centering
    \includegraphics[width=1.0\textwidth]{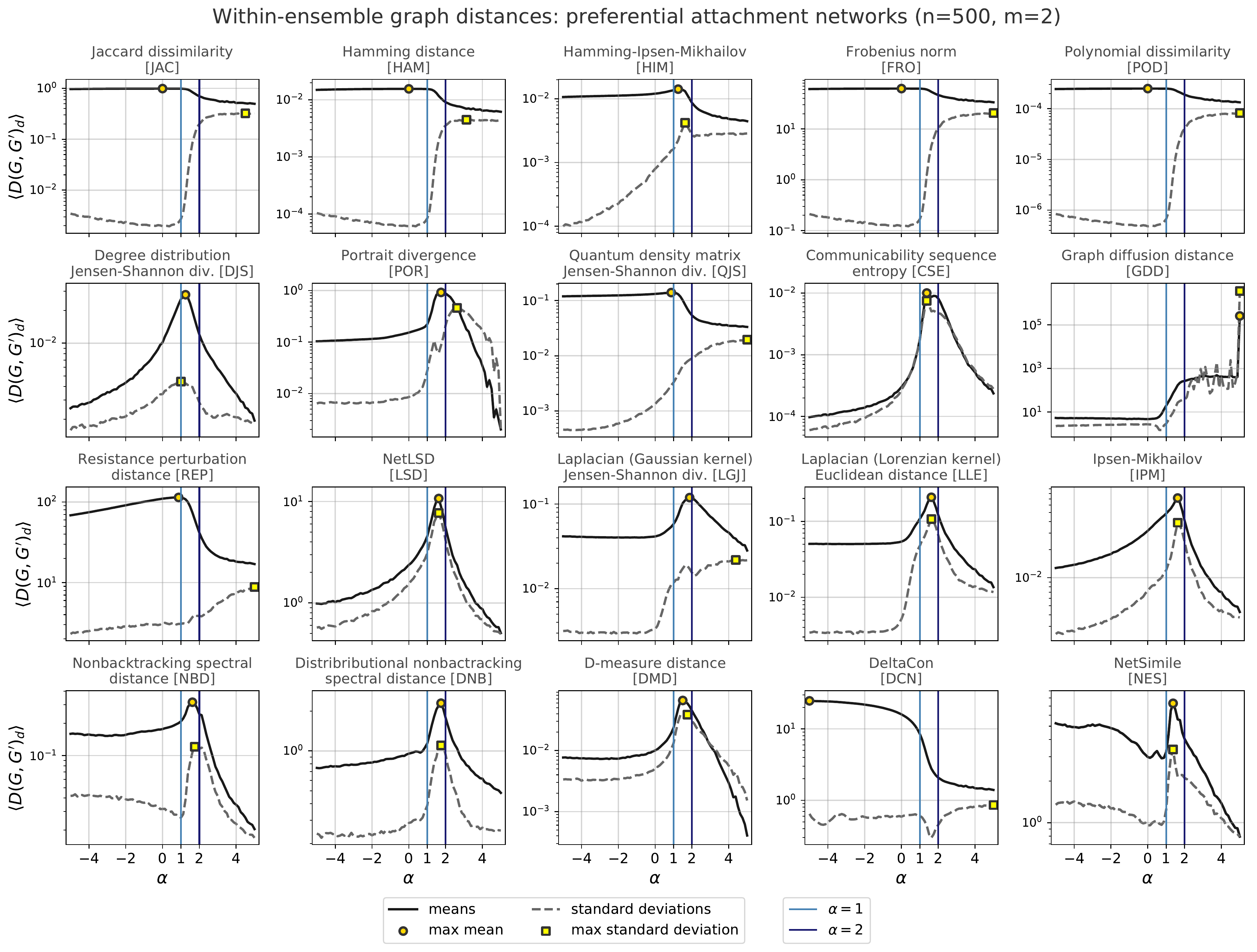}
    \caption{\textbf{Mean and standard deviations of the within-ensemble distances for preferential attachment networks.} Here, we generate pairs of preferential attachment networks, varying the preferential attachment kernel, $\alpha$, while keeping the size and average degree constant. As $\alpha \rightarrow \infty$, the networks become more and more star-like, and at $\alpha =1$, this model generates networks with power-law degree distributions. The mean within-ensemble graph distance is plotted as a solid line with a shaded region around for the standard error ($\langle D \rangle \pm \sigma_{\langle D \rangle} $; note that in most subplots  above, the standard error is too small to see), while the dashed lines are the standard deviations.}
    \label{fig:pa_wegd}
\end{figure*}

\subsection{Results for sparse heterogeneous ensembles}

The sparse graph setting is much closer to that of real networks, which often also have heavy-tailed degree distributions \cite{Newman2005}. This motivated the selection of the following two heterogeneous, sparse ensembles.

\subsubsection{Soft configuration model: heavy-tailed degree distribution}

We study these graphs using a (soft) configuration model with a power-law expected degree distribution; i.e., the expected degree $\kappa$ of a node is drawn proportionally to $\kappa^{-\gamma}$. From this model, we expect two important features that graph distance measures could recover:
\begin{enumerate}
    \setcounter{enumi}{6}
    \item For $\gamma < 3$, we know the variance of the degree diverges in the limit of large graph size $n$ \cite{Newman2005}. Since there should be large variations on the degree sequences for two finite instances, we might also expect the graph distances to produce maximal distance $\langle D \rangle$.
    \item We might also expect a \textit{monotonic} decay in the within-ensemble graph distance as $\gamma$ increases. For large $\gamma$, most expected node-degrees will be approximately the average degree, making the network as a whole structurally similar to an $ER$ graph. On the other hand when $\gamma$ is small (especially when $\gamma\le 3$), there is a wide diversity in the degrees of nodes within the graph, and of the expected degrees of nodes across graphs (since expected degrees are i.i.d. sampled from a Pareto distribution).
\end{enumerate}

Out of the 20 studied, most distances capture both of these features. Since $\gamma$ tunes the degree-heterogeneity (larger $\gamma$ yielding more homogeneous graphs), a decrease in the average distance among pairs of graphs might be expected. For large $\gamma$, most expected node-degrees will be approximately the average degree, making the network as a whole structurally similar to an $ER$ graph. On the other hand when $\gamma$ is small (especially when $\gamma\le 3$), there is a wide diversity in the degrees of nodes within the graph, and of the expected degrees of nodes across graphs (since expected degrees are i.i.d. sampled from a Pareto distribution). Thus a reasonable expectation would be that pairs of graphs on average become farther apart as $\gamma$ is decreased. This is observed in many distances, but with the exceptions of \texttt{QJS} and \texttt{REP}, which each instead exhibit maxima at certain finite values of $\gamma>2$. Additionally, several distances (\texttt{HAM}, \texttt{POR}, \texttt{NBD}, and \texttt{NES}) appear to decay monotonically beyond some very small value of $\gamma$, below which they have a slightly smaller value. This fact could have arisen as a finite-size effect or due to some other details of the implementation, since fluctuations become highly pronounced as $\gamma\rightarrow 2$.

Only one graph distance produces completely unexpected behavior: \texttt{DCN} yields $\langle D \rangle$ that monotonically increases with the scale exponent $\gamma$ of the degree distribution, and its standard deviation is \textit{minimized} when $\gamma \approx 3$. We will expand upon this in the following section.

\subsubsection{Nonlinear preferential attachment}

The final ensemble we include here is the nonlinear preferential attachment growth model. By varying the preferential attachment kernel, parameterized by $\alpha$, we can capture a range of network properties:

\begin{enumerate}
    \setcounter{enumi}{8}
    \item As $\alpha \rightarrow -\infty$, this model generates networks with maximized average path lengths, whereby each new node connects its $m$ links to nodes with the smallest average degree; conversely $\alpha \rightarrow \infty$ generates star-like networks \cite{krapivsky2000connectivity}, an effect known as \textit{condensation}.
    \item At $\alpha = 1$, linear preferential attachment, we see the emergence of scale-free networks \cite{Albert1999}, whereas uniform attachment $\alpha=0$ gives each node an equal chance of receiving the incoming node's links.
\end{enumerate}

When $\alpha=1$, this ensemble theoretically generates networks with power-law degree distributions (with degree exponent, $\gamma = 3$ \cite{Albert2002}), which is reminiscent of the results in Figure \ref{fig:scm_wegd} where we measure the within-ensemble graph distances while varying $\gamma$.

Various mean within-ensemble distances are maximized in the range $\alpha\in[1,2]$, which is indicative of the diversity of possible graphs that can be produced by the preferential attachment mechanism in the small-$\alpha$ regime. For $\alpha\ll 0$, newly arriving nodes connect primarily to the lowest-degree existing nodes (for example leading to long chains of degree-$2$ nodes when $m=1$), making many distance measures record i.i.d. pairs of graphs as similar. For $\alpha\gg 0$, new nodes tend to connect to the highest-degree existing node, leaving a star-like network---then likewise many graph-pairs are deemed very similar. In the intermediate range (e.g. linear preferential attachment, $\alpha=1$), a much wider variety of possible graphs can arise. Thus on average, i.i.d. pairs are (usually) measured as farthest apart in that range.

For preferential attachment networks, we again see curious behavior for \texttt{DCN} where, unlike most other distance measures, heterogeneous graphs with $1 \leq \alpha < 2$ have smaller within-ensemble graph distances than more homogeneous graphs $\alpha < 0$. Upon closer examination, we know why this happens, and to conclude this section, we will walk through the anatomy of \texttt{DCN} and show why its behavior is often different than the other distance measures studied here, especially for heterogeneous networks.

The \textit{descriptor}, $\psi_G$ that \texttt{DCN} is based off of is an \textit{affinity matrix} of the graph (constructed from a belief propagation algorithm, see SI \ref{sec:SI_distances_DCN} for full methodology), while the \textit{distance} is calculated using the Matusita distance (similar to the Euclidean distance). The authors note that they selected this distance because they found that it gave more desirable results: ``...it `boosts' the node affinities and, therefore, detects even small changes in the graphs (other distance measures, including [Euclidean distance], suffer from high similarity scores no matter how much the graphs differ)'' \cite{koutra2016deltacon}. What the choice of the Matusita distance has apparently obscured, however, is a greater specificity for distinguishing heterogeneous networks. We know this because of preliminary experiments where the Matusita distance is swapped out for a Jensen-Shannon divergence (as in, for example, \texttt{CSE}); this resulting within-ensemble graph distance \textit{is} maximized for heterogeneous networks ($1 <\alpha < 2$).

Finally, as we note in Section \ref{sec:results_dense}, we are not asserting that a graph distance measure \textit{should} detect the unique behavior of linear preferential attachment ($\alpha = 1$). Nor are we advocating for practitioners to abandon the use of \texttt{DCN}. What we are claiming, however---and why we chose to focus on \texttt{DCN} in this section---is that we need useful benchmarks for understanding the effects of choosing one descriptor-distance pairing over another. Furthermore, this benchmark \textit{should} be based on the within-ensemble graph distances from well-known ensembles.

\section{Discussion}\label{sec:disc}
Graph ensembles are core to the characterization and broader study of networks. Graphs sampled from a given ensemble will highlight certain observable features of the ensemble itself, and in this work, we have used the notion of graph distance to further characterize several commonly studied graph ensembles. The present study focused on one of the simplest quantities to construct given a distance measure and a graph ensemble, namely the mean within-ensemble distance $\langle D \rangle$. Note however that there are many ensembles for which the present methods could be repeated, as well as more graph distance measures, and infinitely many other statistics that could be examined from the within-ensemble distance distribution. Despite examining the within-ensemble graph distances for only five different ensembles, we observed a richness and variety of behaviors among the various distance measures tested. We view this work as the starting point for more inquiries into the relationship between graph ensembles and graph distances.

One promising future direction for the study of within-ensemble graph distances is the prospect of deriving functional forms for various distance measures, as we do for \texttt{JAC}, \texttt{HAM}, and \texttt{FRO} in SI \ref{sec:SI_Jaccard_analytic}, \ref{sec:SI_Hamming_analytic}, and \ref{sec:SI_Frobenius_analytic}. Other distance measures, such as \texttt{DJS}, likely have approximate analytical expressions derived for certain graph ensembles.

We have here only studied the behavior of graphs within a given ensemble and parameterization, which is essentially the simplest possible choice. This leaves wide open any questions regarding distances between graphs sampled from \textit{different} ensembles---or even different from two different parameterizations of the same ensemble. These will be the topic of follow-up works. Nevertheless, such follow-ups will likewise only cover a very small fraction of all possible combinations.

We hope that our approach will provide a foundation for researchers to clarify several aspects of the network comparison problem. First, we expect that practitioners will be able to use the within-ensemble graph distance in order to \textit{rule out} sub-optimal distance measures that do not pick up on meaningful differences between networks in their domain of interest (e.g., what is an informative ``description-distance'' comparison between brain networks may not be as informative when comparing, for example, infection trees in epidemiology). Second, we expect that this work will provide a foundation for researchers looking to develop new graph distance measures (or hybrid distance measures, such as \texttt{HIM}) that are more appropriate for their particular application areas.

There were 20 different graph distances used in this work, with undoubtedly more that we have not included. Each of these measures seek to address the same thing: quantifying the dissimilarity of pairs of networks. We see the current work as an attempt to consolidate all such methods into a coherent framework---namely, casting each distance measure as a mapping of two graphs into a common descriptor space, and the application of a distance measure within that space. Not only that, we also suggest that stochastic, generative, graph models---because of known structural properties and certain critical transition points in their parameter space---are the ideal tool to use for characterizing and benchmarking graph distance measures.

Classic random graph models can fill an important gap by providing well-understood benchmarks on which to test distance measures \textit{before} using them in applications. Much like in other domains of network science, having effective and well-calibrated comparison procedures is vital, especially given the great diversity of graph ensembles under study and of networks in nature.

\section*{Software and data availability}\label{sec:replication}
All the experiments in this paper were conducted using the \texttt{netrd} Python package \url{https://github.com/netsiphd/netrd}. A repository with replication materials can be found at \url{https://github.com/jkbren/wegd}.

\section*{Acknowledgements}\label{sec:ack}
The authors thank Tina Eliassi-Rad, Dima Krioukov, and Leo Torres for helpful comments about this work throughout. This work was supported in part by the Network Science Institute at Northeastern University and the Vermont Complex Systems Center. B.K. acknowledges support from the National Defense Science \& Engineering Graduate Fellowship (NDSEG) Program. G.S and C.M  acknowledge support from the Natural Sciences and Engineering Research Council of Canada and the Sentinel North program, financed by the Canada First Research Excellence Fund. L.H.D. and A.D. acknowledge support from the National Science Foundations Grant No. DMS-1829826.

\section*{Author contributions}\label{sec:contrib}
All authors contributed to the conception of the project. H.H., A.D., \& L.H.D. devised the formalism used in this work. H.H., B.K., and S.M. conducted simulations of the within-ensemble distances. S.M. led the development of the \texttt{netrd} software package that that was used to perform the analyses. All authors contributed to writing the manuscript. H.H. \& B.K. contributed equally.

\bibliography{main} 
\vfill
\clearpage
\appendix

\section{Within-ensemble graph distance as network size increases}\label{sec:wegd_n}

In Figures \ref{fig:gnp_rgg_wegd}, \ref{fig:gnk_wegd}, \ref{fig:ws_wegd}, \ref{fig:scm_wegd}, and \ref{fig:pa_wegd}, we plot the within-ensemble graph distances of networks with a fixed size. However, one important behavior of graph distance measures is how they change as networks increase in size.

As an example, the Jensen-Shannon divergence between the degree distributions (\texttt{DJS}) of two $ER$ graphs will decrease as $n\rightarrow\infty$, since the empirical degree distributions get closer and closer to a binomial distribution. On the other hand, for graph distances that are explicitly accompanied by a size-normalizing term (e.g. \texttt{HAM}), we would expect that the mean within-ensemble graph distance does not change as network size increases.

\begin{figure*}[t!]
    \centering
    \includegraphics[width=1.0\textwidth]{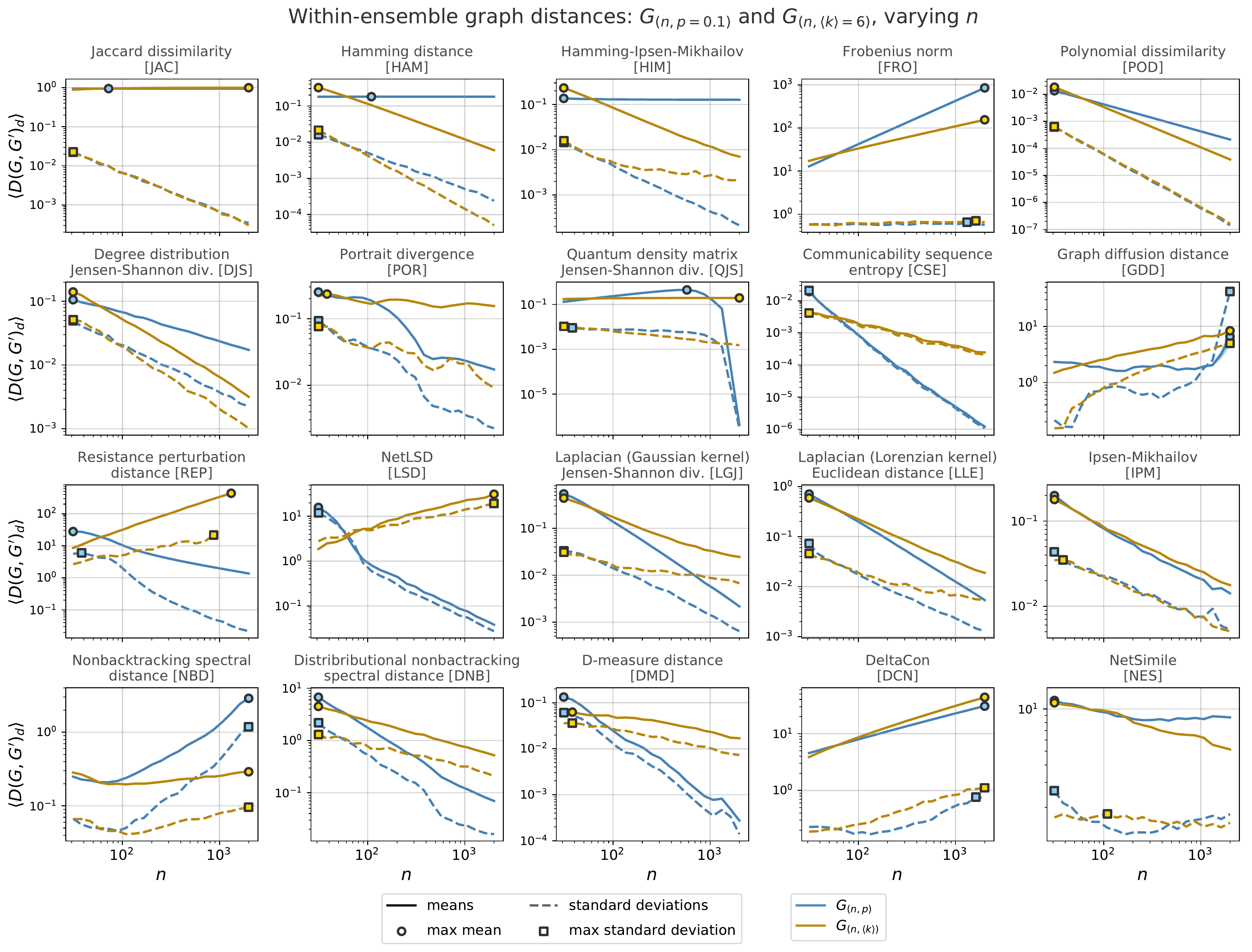}
    \caption{\textbf{Mean and standard deviations of the within-ensemble distances for $G_{(n,p)}$ and $G_{(n,\langle k \rangle)}$ as $n$ increases.} Here, we generate pairs of $ER$ networks with either a fixed density, $p$ or with a fixed average degree, $\langle k \rangle$, as we increase the network size, $n$. In each subplot, the mean within-ensemble graph distance is plotted as a solid line with a shaded region around for the standard error ($\langle D \rangle \pm \sigma_{\langle D \rangle} $; note that in most subplots  above, the standard error is too small to see), while the dashed lines are the standard deviations.}
    \label{fig:n_wegd}
\end{figure*}

In Figure \ref{fig:n_wegd}, we show how the within-ensemble graph distance changes as $n$ increases, both for a fixed density in $G_{(n,p)}$ as well as a fixed average degree in $G_{(n,\langle k \rangle)}$.

\section{Descriptions of graph distance measures}\label{sec:SI_distances}

Throughout the appendix, we assume graphs $G$ and $G'$ are undirected and unweighted so that the adjacency matrices are binary and symmetric. We first consider several projections for distances given a description which is the full adjacency sequence or matrix, followed by projections involving statistical and ad-hoc descriptions. The list of graph distances used in this work is \{\texttt{JAC}, \texttt{HAM}, \texttt{HIM}, \texttt{FRO}, \texttt{POD}, \texttt{DJS}, \texttt{POR}, \texttt{QJS}, \texttt{CSE}, \texttt{GDD}, \texttt{REP}, \texttt{LSD}, \texttt{LGJ}, \texttt{LLE}, \texttt{IPM}, \texttt{NBD}, \texttt{DNB}, \texttt{DMD}, \texttt{DCN}, \texttt{NES}\}.

\subsection{Jaccard Distance}\label{sec:SI_distances_JAC}
The Jaccard measure is computed using the adjacency matrix $\psi_{G} = \mathbf{A}\in\{0,1
\}^{n\times n}$. For two graphs vertex-labeled $G$ and $G'$,
\begin{equation}
    D_{\texttt{JAC}}(G, G') = d_{\texttt{JAC}}(\mathbf{A}, \mathbf{A'}) = 1 - \frac{|\textbf{S}|}{|\textbf{T}|}
\end{equation}
where  $S_{i j} = A_{i j} A^{'}_{i j}$ represents the intersection of edge sets between graphs $G$ and $G'$, while $T_{i j} = S_{i j} + (1-A^{'}_{i j}) A_{i j} + (1-A_{i j})A^{'}_{i j}$ represents the union of edge sets between graphs. Here, $|\textbf{S}|$ is the sum over the $S_{ij}$ and similarly for $|\textbf{T}|$. The computational complexity of the Jaccard distance is $O(|E| + |E'|)$ when using unordered sets to get the union and intersection sets and their cardinality. This is what is done in the \texttt{netrd} package \cite{McCabe2019}.

Since nearly empty graphs likely have nearly zero edges in common, the $|\textbf{S}||\textbf{T}|$ will be nearly zero for $p$ close to $0$, so that $d_{\texttt{JAC}}$ approaches $1$ at low $p$.

\subsection{Hamming Distance}\label{sec:SI_distances_HAM}
Similarly, the Hamming measure may also be computed using the adjacency matrix $\mathbf{A}\in\{0,1\}^{n\times n}$. For two vertex-labeled graphs $G$ and $G'$, the Hamming distance counts the number of elementwise differences between $\psi_G=\mathbf{A}$ and $\psi_{G'}=\mathbf{A}'$:
\begin{equation}
    D_{\texttt{HAM}}(G, G') :=\frac{1}{\binom{n}{2}} \sum_{1\le i<j\le n }|A_{ij} - A'_{ij}|.
\end{equation}
The computational complexity of the Hamming distance is $O(n^2)$ if one compares the elements  $A_{ij}$ for each node-pair $ij$. This is what is used in the \texttt{netrd} package \cite{McCabe2019}. For sparse graphs, one could use unordered sets of edges to only compute the distance on edges $(i,j)$ in the union set $E \cup E'$, leading to a computational complexity of \mbox{$O(|E| + |E'|)$}.

\subsection{Frobenius}\label{sec:SI_distances_FRO}
The Frobenius distance $d_{\texttt{FRO}}$ is simply the norm of matrices, so that:
\begin{equation}
   D_{\texttt{FRO}}(G, G') := \sqrt{\sum_{i,j} |A_{ij} - A'_{ij}|^2}
\end{equation}

Note that for binary adjacency matrices, $|A_{ij}-A'_{ij}|^2=|A_{ij}-A'_{ij}|$, and $A_{ii}=A'_{ii}=0 \; \forall i$ given that there are no self-loops. Note that, because the distance operates on the adjacency matrices directly, it implicitly assumes the graphs are vertex-labeled. \texttt{FRO} has the same computational complexity as the Hamming distance due to their similarity. It is  $O(n^2)$ if one compares all entries, as is in the \texttt{netrd} package, but it could be improved to $O(|E| + |E'|)$.

\subsection{Polynomial Dissimilarity}
\label{sec:SI_distances_POD}

The polynomial dissimilarity, \texttt{POD}, between two unweighted, vertex-labeled graphs is based on the eigenvalue decompositions of the two adjacency matrices of the graphs, $G$ and $G'$ \cite{Donnat2018}.

To compute the polynomial dissimilarity between two graphs, first decompose  $A$ as $Q_A \Lambda_A Q_A^T$, where $Q_A$ is an orthogonal matrix and $\Lambda_A$ is the diagonal matrix of eigenvalues. Second, construct vectors $P(A)$ and $P(A')$ for each graph, where $P(A) = Q_AW_AQ_A^T$ and $W_A=\Lambda_A + \frac{1}{(n-1)^\alpha}\Lambda^2_A + ... + \frac{1}{(n-1)^{\alpha(K-1)}}\Lambda^K_A$.

The polynomial dissimilarity, then, is calculated as the Frobenius norm between $P(A)$ and $P(A')$
\begin{equation}
    D_{\texttt{POD}}(G, G') = \frac{1}{n^2} || P(A) - P(A') || \;.
\end{equation}

In this work, we consider a default value of $K=5$ in order to accommodate potentially informative higher-order interactions in each of the graphs. Here, $\alpha=1$ by default, though in \cite{Donnat2018}, $\alpha=0.9$ is commonly considered.

The computational complexity of \texttt{POD} is $O(n^3)$ in practice, which arises from it requiring two $n\times n$ matrix eigendecompositions, which is $O(n^3)$ for general matrices and a method based on the QR algorithm \cite{Pan1999}, as used in the \texttt{netrd} package. Note that recent techniques based on message-passing can give fast and exact results for sparse networks with short loops in $O(n\log n)$ \cite{Cantwell2019} and could be used to reduce the computational complexity of spectral graph distances.

\subsection{Degree Distribution Jensen-Shannon Divergence}

A simple graph distance measure is the Jensen-Shannon divergence \cite{lin1991divergence} between the empirical degree distributions of two graphs. In this case for an $n$-node graph $G$ the descriptor $\psi_G$ is the empirical degree distribution encoded in the set of numbers $\{p_k(G)\}_{k\ge 0}:=\mathbf{p}$ given by $p_k(G):=n_k(G)/n$, where $n_k(G)=\sum_{i=1}^n\mathbf{1}\{k_i=k\}$, with $\mathbf{1}\{\cdot\}$ being the indicator function and $k_i=\sum_{j=1}^n A_{ij}$ being the degree of node $i$ in terms of the adjacency matrix $\mathbf{A}$ of $G$. The Jensen-Shannon divergence between two such distributions \cite{Carpi2011} is the \textit{degree} Jensen-Shannon divergence or \textbf{DJS} distance between the graphs:
\begin{equation}
    D_{\texttt{DJS}}(G,G') = H\left[\mathbf{p}_+\right] - \dfrac{1}{2}\left(H[\mathbf{p}] + H[\mathbf{p}']\right),
\end{equation}    
where $\mathbf{p}_+=\{(p_k+p_k')/2\}_{k\ge 0}$ is a mixture distribution and $H[\mathbf{p}]=-\sum_{k}p_k\ln p_k$ is the Shannon entropy.

The computational complexity of \texttt{DJS} is $O(n)$, which arises from computing two degree distributions (which is $O(n)$) and then comparing them (which is $O(k_+)$, with $k_+ < n$ being the maximum degree in either network). 

\subsection{Portrait Divergence}\label{sec:SI_distances_POR}

The portrait divergence, \texttt{POR}, compares using the JSD a description for each of two graphs called their \textit{network portrait} \cite{Bagrow2008}. The network portrait is a matrix $B$ with elements $B_{lk}$ such that 
\begin{align}
    B_{lk} \equiv \text{number of nodes with } k \text{ nodes at distance } l.
\end{align}

Alternatively stated, $B_{lk}$ is the $k$th entry of the empirical histogram of $l$-th neighborhood sizes. These elements are computed using a breadth-first search or similar method. The portrait divergence of $G$ and $G'$ is the JSD of probability distributions associated with their portraits, $B$ and $B'$ \cite{Bagrow2019}. Note that each row in $B$ can be interpreted as the probability distribution that there will be $k$ nodes at a distance of $l$ away from a randomly chosen node such that:
\begin{equation}
    P(k|l) = \frac{B_{l,k}}{N} 
\end{equation}
which can be normalized of the number of paths of length $l$ such that the probability distribution is the probability that two randomly selected nodes are at a distance $l$ away from each other:
\begin{equation}
    P(l) = \frac{\sum_{k=0}^n kB_{l,k}}{\sum_c n_c^2}  
\end{equation}
where $n_c$ is the number of nodes within a connected component, $c$. The joint probability of choosing a pair of nodes at a distance, $l$, away from each other \textit{and} that one node has $k$ nodes in total at distance, $l$, away is:
\begin{equation}
    P(k,l) = P(k|l)P(l) = \left( \frac{\sum_{k'=0}^n k'B_{l,k'}}{n} \right) \frac{B_{l,k}}{\sum_c n_c^2} 
\end{equation}

There is now a $P_{B}(k,l)$ and $P_{B'}(k,l)$ for each portrait, $B$ and $B'$, as well as a ``mixed'' distribution for both, which is specified as $P^{*} = \frac{1}{2}(P_{B}(k,l) + P_{B'}(k,l))$. The portrait divergence between $G$ and $G'$ is the JSD between their portraits as follows

\begin{align}
    D_\texttt{POR}(G, G') =& JSD(P_{B}(k,l),P_{B'}(k,l))\nonumber\\= \frac{1}{2} \big(& D_{KL}(P_{B}(k,l) , P^*)+\nonumber\\& D_{KL}(P_{B'}(k,l), P^*) \big)
\end{align}
where $D_{KL}$ is the Kullback-Leibler divergence. Note that $\sqrt{D_{\texttt{POR}}}$ satisfies the properties of a metric (satisfies the triangle inequality, is positive-definite, symmetric) \cite{sutherland2009introduction}.

The computational complexity of \texttt{POR} is $O(n (n + |E|) \log n)$, which comes from the requirement of computing shortest paths between all pairs of nodes in the network. In our implementation, computing the shortest path between a source and all nodes is done with the Dijkstra's algorithm with a binary heap, which takes $O((n + |E|)\log n)$ operations in the worst case. Constructing the portrait and calculating the JSD between the associated distributions has a lower computational complexity.

\subsection{Quantum Spectral Jensen-Shannon Divergence}

This method compares graphs via the Jensen-Shannon divergence (JSD) between probability distributions associated with density matrices of two graphs $G$ and $G'$ \cite{bai2015quantum,rossi2013characterizing,rossi2015measuring, Masuda2019}, denoted $\rho$ and $\rho'$ respectively, defined by
\begin{equation}
    \rho = \dfrac{e^{- \beta \mathbf{L}(G) }}{Z}
\end{equation}
where $\mathbf{L}(G)$ is the Laplacian matrix of graph $G$, and constant $Z \equiv \sum_{i=1}^n e^{- \beta \lambda_i (\mathbf{L})}$, with  $\lambda_i(\mathbf{L})$ being the $i$th eigenvalue of $\mathbf{L}$. Description-distance pair ($\rho$, JSD) yields the ``Quantum Spectral Jensen-Shannon Divergence'' (\texttt{QJS}) \cite{DeDomenico2016SpectralComparison}, which compares two graphs by the entropy of the eigenvalue spectra of their density matrices $\rho$. Treating the spectrum $\{\lambda_i\}_{i=1}^n$ as a normalized probability distribution, the spectral Rényi entropy of order $q$ is given by
\begin{equation}
    S_q = \frac{1}{1-q} \log_2 \sum_{i=1}^{n}\lambda_i(\rho)^q,
\end{equation}
which, if $q=1$, reduces to the Von Neumann entropy:
\begin{equation}
    S_1 = - \sum_{i=1}^{n}\lambda_i(\rho)\log_2\lambda_i(\rho).
\end{equation} 

The QJS distance between two graphs is defined to be:
\begin{equation}
    D_{\texttt{QJS}}(G, G') = S_q\left( \frac{\rho + \rho'}{2} \right) - \frac{1}{2} [S_q(\rho) + S_q(\rho')].
\end{equation}

For default parameter values, we use $\beta=0.1$ and $q=1.0$, based on the explanations in \cite{DeDomenico2016SpectralComparison}. \texttt{QJS} requires computation of Laplacian matrix spectra of two graphs, and comparison thereof, which yields a computational complexity of $O(n^3)$ (see Appendix \ref{sec:SI_distances_POD}).

\subsection{Communicability Sequence Entropy Divergence}\label{sec:SI_distances_CSE}
The communicability sequence entropy divergence $\texttt{CSE}$ between two graphs, $G$ and $G'$, is the JSD between the communicability distributions of $G$ and $G'$. In order to have a communicability distribution, we first construct the communicability matrix, which is an $n\times n$ matrix corresponding to the \textit{communicability} between two nodes, $v_i$ and $v_j$. 
\begin{equation}
    C = e^A = \sum_{k=0}^\infty \frac{1}{k!}A^k
\end{equation}

In other words, the communicability matrix, $C$, is computed as a matrix exponentiation of the adjacency matrix. The elements $C_{ij} , i\leq j,$ are stored in a vector (of length $\binom{n}{2}$) and normalized to create the communicability sequence, $P$ and $P'$, for each graph. The Shannon entropy of $P$ is $H[P] = -\sum_{i=1}^M P_i\log_2P_i$, and the communicability sequence entropy divergence is calculated as the JSD between $P$ and $P'$, where $M$ is the mixed sequence of $P$ and $P'$.
\begin{equation}
    D_\texttt{CSE}(G, G') = JSD(P, P') = H[M] - \dfrac{1}{2}(H[P] + H[P']) \;.
\end{equation}   

The computational complexity of \texttt{CSE} is $O(n^3)$, with the computationally intensive step being to compute the exponential of both adjacency matrices $A$ and $A'$. Our implementation uses Padé approximants through the SciPy package to perform this step, which takes $O(n^3)$ operations to get an approximation \cite{Moler2003}.

\subsection{Graph Diffusion Distance}\label{sec:SI_distances_GDD}

The graph diffusion distance \cite{Hammond2013} $\texttt{GDD}$ between two graphs, $G$ and $G'$, is a distance measure based on the notion of \textit{flow} within each graph. As such, this measure uses the unnormalized Laplacian matrices of both graphs, $\mathbf{L}$ and $\mathbf{L}'$, and uses them to construct time-varying Laplacian exponential diffusion kernels, $e^{-t\mathbf{L}}$ and $e^{-t\mathbf{L}'}$, by effectively simulating a diffusion process for $t$ timesteps (as a default, $t=1000$), creating a column vector of node-level activity at each timestep.

The distance $d_\texttt{GDD}(G, G')$ is defined as the Frobenius norm between the two diffusion kernels at the timestep $t^{*}$ where the two kernels are maximally different.
\begin{equation}
    D_{\texttt{GDD}}(G, G') = \sqrt{||e^{-t^{*}\mathbf{L}} - e^{-t^{*}\mathbf{L}'}||}
\end{equation}

The computational complexity is $O(n^3)$ since a spectral decomposition of the Laplacian matrices is used (see Appendix \ref{sec:SI_distances_POD}).

\subsection{Resistance Perturbation Distance}\label{sec:SI_distances_REP}
The resistance perturbation distance $\texttt{RES}$ between two vertex-labeled graphs, $G$ and $G'$, is the $p$-norm of the difference between two graph resistance matrices \cite{Monnig2018}. The resistance perturbation distance changes if either graph is relabeled (it is not invariant under graph isomorphism), so node labels should be consistent between the two graphs being compared. The distance is not normalized.

The resistance matrix of a graph $G$ is calculated as
\begin{equation}
    R = \text{diag}(\mathcal{L}) \mathbf{1}^T + \mathbf{1} \text{diag}(\mathcal{L})^T - 2\mathcal{L},
\end{equation}
where $\mathcal{L}$ is the Moore-Penrose pseudoinverse of the Laplacian of $G$.

The resistance perturbation graph distance of $G$ and $G'$ is calculated as the $p$-norm (the $p$th root of the sum of the $p$th powers of elements) of the difference in their resistance matrices, $R^{(1)}$ and $R^{(2)}$
\begin{align}
     D_{\texttt{REP}}(G, G') &= \left[\sum_{i,j \in V} |R_{i,j} - R'_{i,j}|^p\right]^{1/p}.
\end{align}
The default value chosen in experiments is $p=2$. The computational complexity of $\texttt{RES}$ is $O(n^3)$ for our implementation, since we need to compute the Moore-Penrose pseudoinverse of the Laplacian matrix of both graphs, which is $O(n^3)$. Note that low-rank approximations can be used to reduce the computational complexity \cite{Monnig2018}.

\subsection{NetLSD}\label{sec:SI_distances_LSD}

The NetLSD distance $\texttt{LSD}$ between two graphs, $G$ and $G'$, is the Frobenius norm between the heat trace signatures of the normalized Laplacians $\mathbf{L}$ and $\mathbf{L}'$ \cite{Tsitsulin2018}. The heat kernel matrix is calculated as
\begin{equation}
    H_t = e^{-t \mathbf{L}} = \sum_{j=1}^n e^{-t \lambda_j} \phi_j \phi_j^T.
\end{equation}
    The $ij$-th element of $H_t$ contains the amount of heat transferred from node $v_i$ to node $v_j$ at time $t$ (default of 256 log-spaced time intervals between $10^{-2}$ and $10^{2}$). From the heat kernel matrix $H_t$, the \textit{heat trace}, $h_t$ is defined as
\begin{equation}
    h_t = \text{Tr}(H_t) = \sum_{j=1}^n e^{-t \lambda_j}.
\end{equation}
The {\it heat trace signature} of graph $G$ is the set $\{h_t\}_{t\ge 1}$. Upon computing heat trace signatures of both $G$ and $G'$, they are compared via a Frobenius norm
\begin{equation}
    D_\texttt{LSD}(G,G') = d_\texttt{FRO}\left(\{h_t\}_{t\ge 0},\{h'_t\}_{t\ge 0}\right).
\end{equation}

The computational complexity of \texttt{LSD} is $O(n^3)$ due to the spectral decomposition of the Laplacian matrices of both graphs (see Appendix \ref{sec:SI_distances_POD}).

\subsection{Laplacian Spectrum Distances}\label{sec:SI_distances_LS}

Many distances between two graphs, $G$ and $G'$, use a direct comparison of their Laplacian spectrum. For all the methods below, we use the eigenvalues $\{ \lambda_1 = 0 \leq \lambda_2 \leq \dots \leq \lambda_{n} \}$ of the normalized Laplacian matrices $\mathbf{L}$ and $\mathbf{L}'$. To perform the comparison, a subset of the whole spectrum can be used, e.g. the $k$ smallest \cite{Wills2020} or largest \cite{Jurman2011,Masuda2019} in magnitude. Unless specified, we used all eigenvalues for comparison ($k = n$).

The distances compare the continuous spectra $\rho(\lambda)$ and $\rho'(\lambda)$ associated with the graph $G$ and $G'$. A continuous spectrum is obtained by the convolution of the discrete spectrum $\sum_{i} \delta(\lambda - \lambda_i)$ with a kernel $g(\lambda, \lambda^*)$
\begin{equation}
    \rho(\lambda) = \frac{1}{Z}\sum_{i = 1}^n \int_0^2 g(\lambda, \lambda^*) \delta(\lambda^* - \lambda_i) \mathrm{d} \lambda^* \;,
\end{equation}
where $Z$ is a normalization factor. Different types of distribution can be used for the kernel, for instance a Lorentzian distribution \cite{Ipsen2002}
\begin{equation}
    g(\lambda, \lambda^*) = \frac{\gamma}{\pi [\gamma^2 + (\lambda-\lambda^*)^2]} \;,
\end{equation}
or a Normal distribution
\begin{equation}
    g(\lambda, \lambda^*) = \frac{\exp[-(\lambda-\lambda^*)^2/2\sigma^2]}{\sqrt{2 \pi \sigma^2}} \;.
\end{equation}

Different types of metrics can then be used to compare the spectra, such as the Euclidean metric
\begin{equation}
    d(\rho,\rho') = \sqrt{\int_0^2 [\rho(\lambda) - \rho'(\lambda)]^2 \mathrm{d}\lambda} \;,
\end{equation}
or the square root of the JSD $d(\rho, \rho') = \sqrt{JSD(\rho, \rho')}$, written as 
\begin{equation}
    JSD(\rho, \rho') = \frac{1}{2}D_{KL}(\rho || \bar{\rho} ) + \frac{1}{2} D_{KL}(\rho' || \bar{\rho})
\end{equation}
where $\bar{\rho} = (\rho + \rho')/2$. Various combination of kernels and metrics yield the following distinct distance measures:
\begin{itemize}
    \item Laplacian spectrum: Gaussian kernel, JSD distance $\texttt{LGJ}$
    \item Laplacian spectrum: Lorenzian kernel, Euclidean distance $\texttt{LLE}$
\end{itemize}
For both kernels, we use a \textit{half width at half maximum} of $0.011775$ (which means the standard deviation for the Gaussian kernel is $\approx 0.01$).

While we only focus on the two specific distances above, we note again that there is a world of possible combinations of descriptor-distance pairs to possibly use for comparing graphs. We selected the two above because their within-ensemble graph distance curves differed the most (e.g. as opposed to including Gaussian kernel / Euclidean distance or Lorenzian kernel / JSD). The computational complexity of this suite of graph distances is $O(n^3)$ due to the spectral decomposition of the Laplacian matrices of both graphs (see Appendix \ref{sec:SI_distances_POD}).

\subsection{Ipsen-Mikhailov}\label{sec:SI_distances_IPM}
The Ipsen-Mikhailov distance \cite{Ipsen2002}  $\texttt{IPM}$ between two graphs, $G$ and $G'$, is a spectral comparison of their Laplacian matrices, $\mathbf{L}$ and $\mathbf{L}'$. This approach treats the set of nodes in $G$ and $G'$ as molecules with an elastic connection between them, which casts the distance measurement between $G$ and $G'$ as the solution to a set of differential equations between the vibrational frequencies between the nodes. The vibrational frequencies, $\omega_i$, of each node in $G$ is related to the eigenvalues, $\lambda$, of $\mathbf{L}$ such that $\lambda_i = \omega_i^2$.

With this, one can construct a spectral density for each graph as a sum of Lorenz distributions as follows
\begin{equation}
    \rho(\omega) = \frac{1}{Z} \sum_{i=1}^{n-1} \frac{\gamma}{(\omega - \omega_i)^2+\gamma^2}
\end{equation}
where $Z$ is a normalization term, and $\gamma$ is a fixed scaling term that controls the width of the Lorenz distributions (as in \cite{Ipsen2002}, we use $\gamma=0.08$ as a default). The distance between $G$ and $G'$ is then calculated as 
\begin{equation}
    D_\texttt{IPM}(G, G') = d(\rho, \rho') = \sqrt{\int_{0}^{\infty} [\rho(\omega) - \rho'(\omega)]^2 d\omega}
\end{equation}    

The computational complexity of \texttt{IPM} is $O(n^3)$ due to the spectral decomposition of the Laplacian matrices of both graphs (see Appendix \ref{sec:SI_distances_POD}).

\subsection{Hamming-Ipsen-Mikhailov}\label{sec:SI_distances_HIM}
The Hamming-Ipsen-Mikhailov distance $\texttt{HIM}$ between two vertex-labeled graphs, $G$ and $G'$ is expressed as a weighted combination of the \texttt{IPM} (Section \ref{sec:SI_distances_IPM}) distance and a normalized \texttt{HAM} (Section \ref{sec:SI_distances_HAM}) distance \cite{Jurman2015}. The parameter $ \gamma$ for the $\texttt{IPM}$ is fixed such that $D_\texttt{IPM}(\mathcal{E}_n, \mathcal{F}_n) = 1$, where $\mathcal{E}_n$ and $\mathcal{F}_n$ are the empty and complete graphs of $n$ nodes. The \texttt{HIM} distance is defined as follows

\begin{equation}
    D_\texttt{HIM}(G,G') = \frac{1}{\sqrt{1+\xi}} \sqrt{D_\texttt{IPM}(G, G')^2 + \xi D_\texttt{HAM}(G, G')^2}
\end{equation}

We default to $\xi=1$, as in \cite{Jurman2015}. The computational complexity of \texttt{HIM} is $O(n^3)$, with the computationally intensive part being the computation of the \texttt{IPM} distance.

\subsection{Non-backtracking Spectral Distance}\label{sec:SI_distances_NBD}
The non-backtracking spectral distance $\texttt{NBD}$ between two graphs, $G$ and $G'$, is a method that compares the eigenvalues of the non-backtracking matrix of each graph, $\mathcal{B}$ and $\mathcal{B}'$ \cite{Torres2019}. This distance is based on the length spectrum and the set of non-backtracking cycles of a graph (i.e., a closed walk that does not immediately return to the node from which it left) and is calculated as the earth mover's distance ($EMD$) between the eigenvalues of $\mathcal{B}$ and $\mathcal{B}'$. The eigenvalues of $\mathcal{B}$ and $\mathcal{B}'$ are expressed as $\lambda_k = a_k + ib_k$ and $\lambda'_k = a'_k + ib'_k$, respectively, and $EMD(\lambda_{\mathcal{B}}, \lambda_{\mathcal{B}'})$ is the solution to an optimization problem finding the minimum amount of work required to move the coordinates of $\lambda$ to the positions of $\lambda'$.
\begin{equation}
    D_\texttt{NBD}(G, G') = EMD(\lambda_{\mathcal{B}}, \lambda_{\mathcal{B}'}) \;.
\end{equation}

Note that the Ihara determinant formula can be used to obtain the non-backtracking eigenvalues different from $\pm 1$ using a $2n \times 2n$ matrix \cite{Torres2019}.

If one uses the whole non-backtracking spectrums to compute the distance, the computational complexity would be $O(n^3)$ \cite{Torres2019}. Instead of using the whole spectrum of the non-backtracking matrices, for graph $G$ we compute only the $r$ eigenvalues larger in magnitude than $\sqrt{\lambda_1}$, where $\lambda_1$ is the largest eigenvalue of $\mathcal{B}$ \cite{Torres2019}.

The computational complexity of our implementation of \texttt{NBD} is $O(\mathrm{max}(r,r') n^2)$ for general graphs, where $r$ and $r'$ are the number of eigenvalues larger in magnitude than $\sqrt{\lambda_1}$ and $\sqrt{\lambda_1'}$, respectively for graph $G$ and $G'$. To compute these eigenvalues, an implicitly restarted Arnoldi method is used. For sparse graphs the computation is even more efficient.

\subsection{Distributional Non-backtracking Distance}\label{sec:SI_distances_DNB}
Similar to the \texttt{NBD} distance \cite{Mellor2019}, the $\texttt{DNB}$ distance leverages spectral properties of the non-backtracking matrices, $\mathcal{B}$ and $\mathcal{B}'$, of two graphs, $G$ and $G'$, in order to calculate their dissimilarity.

Unlike the \texttt{NBD} distance, the \texttt{DNB} involves a comparison of the (re-scaled) distribution of eigenvalues of $\mathcal{B}$ and $\mathcal{B}'$, which are then compared using either the Euclidean distance or the Chebyshev distance (here, we use the Euclidean distance). We also use the whole spectrum for this distance. Therefore, the computational complexity of \texttt{DNB} is $O(n^3)$ due to the spectral decomposition of the two $2n \times 2n$ matrices (see Appendix \ref{sec:SI_distances_NBD}).

\subsection{\textit{D}-measure Distance}\label{sec:SI_distances_DMD}
The \textit{D}-measure distance \cite{Schieber2017} $\texttt{DMD}$ between two graphs, $G$ and $G'$, involves a combination of three properties from the two graphs to be compared, $G$ and $G'$: the \textit{network node dispersion} ($NND$), \textit{the node distance distribution} ($\mu$), and the $\alpha$-$centrality$ ($\alpha$) for each graph. For a full explanation and justification for each of the components involved in this distance, we refer the reader to the original article \cite{Schieber2017}, but we will briefly summarize it below.

In order to compute the $NND$ of a graph, each node, $v_i$, is assigned a probability vector, $\mathbf{P}_i$, with elements that are the fraction of nodes that are connected to $v_i$ at each distance $j \leq d$, where $d$ is the diameter of the network. The $NND$, then, is defined as
\begin{equation}
    NND(G) = \dfrac{JSD\big(\mathbf{P}_1, \mathbf{P}_2, ..., \mathbf{P}_n\big)}{\log (d+1)}
\end{equation}
where $JSD\big(\mathbf{P}_1, \mathbf{P}_2, ..., \mathbf{P}_n\big)$ is the Jensen-Shannon divergence of each $\mathbf{P}_i$ from the whole network's average node-distance distribution at every distance $j$, which we will denote $\mu_j$. The average $\mu_j$ for all distances $j \leq d$ in a graph, $G$, we will denote $\mu_G$.

The final step before the calculation of the \textit{D}-measure distance is to find the $\alpha$-centrality \cite{bonacich1987power} of each network, $G$ and $G'$, as well as the $\alpha$-centrality of the \textit{complement} of each network, $G^c$ and $G^c{}'$. The $\alpha$-centralities of the original networks are denoted $P_{\alpha G}$ and $P_{\alpha G'}$, while the $\alpha$-centralities of their complements are $P_{\alpha G^c}$ and $P_{\alpha G^c{}'}$.

Ultimately, the \textit{D}-measure distance, $D_{\texttt{DMD}}$, between two graphs is as follows:
\begin{align}
    D&_\texttt{DMD}(G, G') = w_1\sqrt{\dfrac{JSD(\mu_G, \mu_{G'})}{\log(2)}} +\nonumber\\& w_2 \Big| \sqrt{NND(G)} - \sqrt{NND(G')} \Big| +\nonumber\\&\dfrac{w_3}{2} \Bigg(\sqrt{\dfrac{JSD(P_{\alpha G},P_{\alpha G'})}{\log (2)}} + \sqrt{\dfrac{JSD(P_{\alpha G^{c}},P_{\alpha G^{c} {}'})}{\log(2)}} \Bigg)
\end{align}
where $w_1+w_2+w_3$ must equal $1.0$. To calculate the final distance value, we adopt the convention used in \cite{Schieber2017} such that $w_1=0.45$, $w_2=0.45$, $w_3=0.1$.

According to Ref.~\cite{Schieber2017}, the computational complexity of \texttt{DMD} is $O(|E|+n\log n)$. However, one needs to compute all shortest paths between all nodes, which suggest a more computationally intensive calculation. We rather have a computational complexity of $O(n(n+|E|) \log n)$ with our implementation using Dijkstra algorithm with a binary heap (see Appendix \ref{sec:SI_distances_POR}).

\subsection{DeltaCon}\label{sec:SI_distances_DCN}
The DeltaCon distance $\texttt{DCN}$ between two graphs, $G$ and $G'$, is the Matusita distance between the \textit{affinity} matrices, $S$ and $S'$, of $G$ and $G'$. The affinity matrices are constructed using Fast Belief Propagation, which is expressed as
\begin{equation}
    [\mathbf{I} + \epsilon^2\mathbf{D} - \epsilon \mathbf{A}]\vec{s_i} = \vec{e_i}
\end{equation}
where $\mathbf{I}$ is the $n\times n$ identity matrix, $\mathbf{D}$ is the diagonal degree matrix, $\mathbf{A}$ is the adjacency matrix, $\vec{e_i}$ is a vector indicating the initial node $v_i$ from which a random walk process is initiated, and $\vec{s_i}$ is a column vector consisting of $s_{ij}$, which is the affinity of node $v_j$ with respect to node $v_i$. The affinity matrices, $S$ and $S'$, are defined as $S=[\mathbf{I} + \epsilon^2\mathbf{D} - \epsilon \mathbf{A}]^{-1}$. The distance between $G$ and $G'$ according to DeltaCon is as follows
\begin{equation}
    D_\texttt{DCN}(G, G') = d(S, S') = \sqrt{\sum_{i=1}^n \sum_{j=1}^n \big(\sqrt{s_{ij}} -  \sqrt{s'_{ij}}\big)^2}
\end{equation}

The computational complexity of our implementation of \texttt{DCN} is $O(n^3)$ since we obtain $S$ by matrix inversion directly. However, note that it is possible to improve the algorithm and have an $O(n^2)$ computational complexity using a power method or even $O(|E|)$ by approximating the distance \cite{koutra2016deltacon}.

\subsection{NetSimile}\label{sec:SI_distances_NSE}
NetSimile $\texttt{NES}$ is a method for comparing two graphs, $G$ and $G'$, that is based on statistical features of the two graphs. It is invariant to graph labels and is able to compare graphs of different sizes \cite{berlingerio2012netsimile}. It is calculated as the Canberera distance between the $7 \times 5$ feature matrix, $\mathbf{p}$ and $\mathbf{p'}$, of each graph. To construct the $\mathbf{p}$ and $\mathbf{p'}$ feature matrices, first a $7 \times n$ matrix is constructed for each, with each column, $j$, consisting of the following seven node-level quantities:
\begin{enumerate}
    \item degree, $k_j=\sum_{j}A_{ij}$
    \item clustering coefficient, $c_j=(A^3)_{jj}/\binom{k_j}{2}$
    \item average neighbor degree $k^{(nn)}_j=\frac{1}{k_j}\sum_{i}k_i A_{ij}$.
    \item average clustering coefficient of the nodes in the ego network $c^{(ego)}_j=\sum_{i}c_i A_{ij}$
    \item number of edges within the ego network $T_j=\sum_{l,m}A_{jl}A_{lm}A_{mj}$
    \item number of outgoing edges from the ego network $O_j=\sum_{i}A_{ij}k_i - T_j = k_j k^{(nn)}_j - T_j$
    \item number of neighbors of the ego network $nn_j^{(ego)}=\sum_{i}\mathbf{1}_{\{\exists l\in \mathcal{N}_j: i\sim l, i\not\sim j\}}$ 
\end{enumerate}

These features are then summarized into $\mathbf{p}$ and $\mathbf{p'}$, which are $7 \times 5$ signature vectors consisting of the median, mean, standard deviation, skewness, and kurtosis of each feature. NetSimile uses the Canberra distance to arrive at a final scalar distance.
\begin{equation}
    D_\texttt{NSE}(G, G') = d(\mathbf{p}, \mathbf{p'}) = \sum_{i=1}^n \frac{|p_i-p'_i|}{|p_i|+|p'_i|}
\end{equation}

The computational complexity of \texttt{NES} depends on two parts : features extraction and features aggregation. Features are all locally defined, hence their extraction will take $O(q n )$ where $q$ is the average degree of a node when selecting a random edge and choosing an endpoint \cite{Henderson2011}. Feature aggregation is $O(n\ln n)$ \cite{berlingerio2012netsimile}, hence the overall complexity is $O(q n + n \log n)$.

\section{Analytical derivation of within-ensemble graph distances}\label{sec:SI_distance_derivations}

\subsection{Jaccard Distance}\label{sec:SI_Jaccard_analytic}
We can directly calculate $\langle d_{\texttt{JAC}}(\mathbf{A}, \mathbf{A'})\rangle_{G(n,p)}$, the expected Jaccard distance among two graphs sampled from $G_{(n,p)}$. Both $|\textbf{T}|$ and $|\textbf{S}|$ are distributed binomially, as they are the sum of $\binom{n}{2}$ Bernoulli values arising with  probability $p^2$ and $2p(1-p)+p^2$, respectively. Since binomial distributions are sharply peaked (for large values of $n$), we can approximate the expected value of the ratio $|\mathbf{S}|/|\mathbf{T}|$ by the ratio of the expected values of $|\mathbf{S}|$ and $|\mathbf{T}|$. Thus we have,

\begin{align}
    \langle d_{\texttt{JAC}}(\mathbf{A}, \mathbf{A'}) \rangle _{G(n,p)} &=1- \left\langle\frac{|\textbf{S}|}{|\textbf{T}|} \right\rangle \nonumber\\&\approx 1-\frac{\langle|\textbf{S}|\rangle}{\langle|\textbf{T}|\rangle} \nonumber\\&=1-\dfrac{p^2\binom{n}{2}}{\left(2p(1-p)+p^2\right)\binom{n}{2}}\nonumber\\&= \frac{1-p}{1-\frac{p}{2}}
\end{align}
which agrees precisely with simulations. Note, in the limit $p\approx 1$, we have by Taylor expansion,
\begin{widetext}
\begin{equation} 
\begin{aligned}
   \langle d_{\texttt{JAC}}(\mathbf{A}, \mathbf{A'})\rangle_{G(n,p\approx 1)} &=1-\left.\left\langle\frac{|\textbf{S}|}{ |\textbf{T}|}\right\rangle \right\vert_{p\approx 1}\\
   &= \left.\frac{1-p}{1-\frac{p}{2}}\,\right\vert_{p=1}+(p-1)\frac{d}{dp}\left. \left( \frac{1-p}{1-\frac{p}{2}} \right)\right\vert_{p=1}+...\\
    &= 0 + (p-1) \left.\left(\frac{-1}{1-\frac{p}{2}}+\frac{-(1-p)(-\frac{1}{2})}{(1-\frac{p}{2})^2}\right)\right\vert_{p=1}+...\\
    &=(p-1)\left(\frac{-1}{1-\frac{1}{2}}+0\right)+...\\
    &=2(1-p)+ ...,
\end{aligned}
\end{equation}
\end{widetext}

Similarly---as we show in SI \ref{sec:SI_Hamming_analytic}---the Hamming distance ($d_{\texttt{HAM}}$) behaves in this region as

\begin{widetext}
\begin{equation} 
\begin{aligned}
    \langle d_{\texttt{HAM}}(\mathbf{A}, \mathbf{A'})\rangle_{G(n,p\approx 1)} &= 2p(1-p) |_{p=1}+(p-1) \left(2(1-p)-2p\right)|_{p=1}+ ...\\
    &=0+(p-1)(0-2)+ ...\\ 
    &=2(1-p)+...,
\end{aligned}
\end{equation}
\end{widetext}
which is exactly the same. Indeed, we observe this equivalence in Figure \ref{fig:gnp_rgg_wegd} in the region $p\approx 1$. This finding makes intuitive sense because in the region $p\approx 1$, the ``union graph'', $\mathbf{T}$, is likely an essentially complete graph, and $d_{\texttt{JAC}}$ simply measures the fraction of edges/non-edges that are not in agreement between $G$ and $G'$, which is precisely what $d_{\texttt{HAM}}$ does for all $p$ given an adjacency description.

\subsection{Hamming Distance}\label{sec:SI_Hamming_analytic}

The Hamming measure is simply the fraction of mismatched entries between $A$ and $A'$. Due to this simplicity, we again can analytically predict the mean within-ensemble graph distance for graphs sampled from $G_{(n,p)}$:

\begin{align}
   \langle d_{\texttt{HAM}}(\mathbf{A}, \mathbf{A'})\rangle_{G(n,p)}&=\frac{1}{\binom{n}{2}} \sum_{1\le i<j\le n } \mathbb{P}(|A_{ij} - A'_{ij}|= 1)\label{eq:HAM}\\&= 2 p(1-p).\nonumber
\end{align}

The function $2p(1-p)$ is $n$-independent, and has a maximum at $p=\frac{1}{2}$; simulations are matched by it precisely. Interestingly, while this calculation was done for $G_{(n,p)}$, the results in Figure \ref{fig:gnp_rgg_wegd} shows an equivalent result for $RGG$s of the same edge-density $p$.

\subsection{Frobenius}\label{sec:SI_Frobenius_analytic}

As a back-of-the envelope-calculation, note that the sum of elementwise differences is binomially distributed with mean $\langle \sum_{i,j}|A_{ij}-A_{ij}'|\rangle=n(n-1)2p(1-p)$. Using sharply-peakedness, we can thus state approximately,

\begin{align}
     \langle d_{\texttt{FRO}}(\mathbf{A}, \mathbf{A'})\rangle_{G(n,p)} &=\left\langle  \sqrt{\sum_{i,j} |A_{ij} - A'_{ij}|^2}\right\rangle \nonumber\\&\approx \sqrt{\left\langle  \sum_{i,j} |A_{ij} - A'_{ij}|\right\rangle}\nonumber\\&\simeq n\sqrt{2p(1-p)}, 
\end{align}
which exhibits a maximum at $p=\frac{1}{2}$ for any given $n$, but grows linearly with $n$, the latter two observations are qualitatively born out in simulations.

\end{document}